\documentclass[aps, pra, amsmath, showpacs, preprintnumbers, groupedaddress, twocolumn, amssymb, a4paper]{revtex4-1}
\usepackage{graphicx,xspace}
\usepackage{epstopdf}
\usepackage{mathptmx, textcomp}
\usepackage[latin1]{inputenc}
\usepackage[T1]{fontenc}
\usepackage{array, multirow}
\usepackage[table]{xcolor}
\usepackage{booktabs}
\usepackage{longtable}

\newcolumntype{L}[1]{>{\raggedright\let\newline\\\arraybackslash\hspace{0pt}}m{#1}}
\newcolumntype{C}[1]{>{\centering\let\newline\\\arraybackslash\hspace{0pt}}m{#1}}
\newcolumntype{R}[1]{>{\raggedleft\let\newline\\\arraybackslash\hspace{0pt}}m{#1}}

\newcommand{\Xstate}[1][+]{\ensuremath{X\:^1\Sigma^{#1}_g}\xspace}
\newcommand{\cstate}[1][+]{\ensuremath{c\:^3\Sigma^{#1}_g}\xspace}
\newcommand{\Astate}[1][+]{\ensuremath{A\:^1\Sigma^{#1}_u}\xspace}
\newcommand{\astate}[1][+]{\ensuremath{a\:^3\Sigma^{#1}_u}\xspace}

\newcommand{\oneG}{\ensuremath{1_{g}}\xspace}
\newcommand{\zeroG}{\ensuremath{0^{-}_{g}}\xspace}
\newcommand{\vib}{\ensuremath{\text{v}}}

\newcommand{\wn}{$\mathrm{cm}^{-1}$}

\definecolor{mark}{rgb}{0.6863,0.9333,0.9333}
\definecolor{lGrey}{RGB}{0.1,0.1,0.1}

\begin{document}
\graphicspath{{plots/}}

\title{Level structure of deeply bound levels of  the \cstate state of $^{87}\text{Rb}_2$ }

\author{Bj\"orn Drews$^1$}
\author{Markus Dei{\ss}$^1$}
\author{Joschka Wolf$^1$}
\author{Eberhard Tiemann$^2$}
\author{Johannes Hecker Denschlag$^1$}
\affiliation{$^1$Institut f\"ur Quantenmaterie and Center for Integrated Quantum Science and Technology IQ$^{ST}$, Universit\"at Ulm, 89069 Ulm, Germany\\$^2$Institut f\"ur Quantenoptik, Leibniz Universit\"at Hannover, 30167 Hannover, Germany}

\date{\today}

\begin{abstract}
We spectroscopically investigate the hyperfine, rotational and Zeeman structure of the vibrational levels $\vib'=0$, 7, 13 within the electronically excited \cstate state of $^{87}\text{Rb}_2$ for magnetic fields of up to $1000\,\text{G}$. As spectroscopic methods we use short-range photoassociation of ultracold Rb atoms as well as photoexcitation of ultracold molecules which have been previously prepared in several well-defined quantum states of the \astate potential. As a byproduct, we present optical two-photon transfer of weakly bound Feshbach molecules into \astate, $\vib=0$ levels featuring different nuclear spin quantum numbers. A simple model reproduces well the molecular level structures of the \cstate vibrational states  and provides a consistent assignment of the measured resonance lines. Furthermore, the model can be used to predict the relative transition strengths of the lines. From fits to the data we extract for each vibrational level the rotational constant, the effective spin-spin interaction constant, as well as the Fermi contact parameter and (for the first time) the anisotropic hyperfine constant. In an alternative approach, we perform coupled-channel calculations where we fit the relevant potential energy curves, spin-orbit interactions and hyperfine functions. The calculations reproduce the measured hyperfine level term frequencies with an average uncertainty of $\pm9\:$MHz, similar as for the simple model. From these fits we obtain a section of the potential energy curve for the  \cstate~state which can be used for predicting the level structure for the vibrational manifold $\vib'=0$ to 13 of this electronic state.
\end{abstract}

\narrowtext

\maketitle

\section{INTRODUCTION}
\label{sec:introduction}

In recent years there has been a renewed interest in spectroscopy of alkali dimers, partially driven by the prospects for experiments with ultracold molecules \cite{Carr2009, Krems2008, Chin2009, DeMille2002}. For such experiments, detailed knowledge of the molecular hyperfine structure is crucial, especially when molecules are required to be in precisely defined quantum states. For the preparation of such molecules in the singlet and triplet ground state often excited electronic states are involved as e.g. in photoassociation \cite{Deiglmayr2008, Aikawa2010, Sage2005, Viteau2008, Bellos2011} or in stimulated Raman adiabatic passage transfer (STIRAP) of Feshbach molecules \cite{Lang2008, Ni2008, Danzl2010, Molony2014, Takekoshi2014, Guo2016, Park2015}. Here, the rotational and hyperfine structure of the intermediate, electronically excited states can be conveniently used to tailor the quantum numbers of the desired ground state molecules (see e.g. \cite{Deiss2014}). While deeply bound \Xstate molecules are quite generally formed with the help of the \Astate state (e.g. \cite{Danzl2010, Molony2014, Takekoshi2014}), \astate molecules are in general produced via the \cstate state (e.g.~\cite{Lozeille2006, Lang2008}) which is the energetically lowest excited state with $g$-symmetry.

In this article, we experimentally investigate the hyperfine structure of the vibrational levels $\vib' =  0$, 7, 13 of the \cstate state of $^{87}$Rb$_2$, extending our previous work on deeply bound levels \cite{Takekoshi2011}. This is complemented by recent work of Tsai \emph{et al.}
\cite{Tsai2013} who investigated the hyperfine structure of the weakly bound levels of \cstate, \oneG.

One important result is a partial change of our previous assignment of the spectra of measured  transition lines towards \cstate, resolving an inconsistency that was already mentioned in \cite{Takekoshi2011} in connection with their Fig.~7. In contrast to the assumption in Ref.~\cite{Takekoshi2011} that in the applied spectroscopy only final levels with a total nuclear spin of $I' = 3$ can be observed we now additionally identify final levels with $I' = 1$. The precision and resolution of our data has considerably improved, and we can describe the data well with a model Hamiltonian from which we obtain for the first time also anisotropic hyperfine parameters $c$ for the state \cstate. This parameter turns out to be much larger than theoretically expected \cite{Lysebo2013}. We experimentally determine the molecular parameters for the vibrational levels $\vib' = 0$, 7, 13 and observe how they change with $\vib'$. A comparison of calculated relative transition strengths to those of the corresponding observations shows very good agreement and proofs the consistent description of the hyperfine structure within the model.

The present work is a compilation of measurements that we have taken over the last three years. These measurements were carried out on two different $^{87}$Rb cold atom set-ups \cite{Deiss2014, Schmid2012} and using two different spectroscopic methods. Method 1 is short-range photoassociation where we expose an optically trapped gas of Rb atoms at a temperature of about $1\,\mu$K to a tunable narrow-linewidth photoassociation laser and measure the atomic losses after a given interaction time. In Method 2 a tunable narrow-linewidth laser photoexcites $^{87}$Rb$_2$ molecules in well defined quantum states (either Feshbach molecules or \astate, $\vib = 0$ triplet molecules) to the \cstate state and inflicts losses which are then measured for a given interaction time.

This article is organized as follows. In section~\ref{sec:expSetup} we give an overview of the used spectroscopic methods followed by a discussion of the obtained spectroscopic data in section~\ref{sec:groundstate}. Then, in section \ref{sec:modelCalc}, our simple theoretical model to analyze the measurements is described and the extracted results are discussed. Finally, section~\ref{sec:ccModel} is dedicated to a presentation of full coupled-channel calculations of the \cstate state imbedded in $^1\Sigma^+_g$, $^1\Pi_g$ and $^3\Pi_g$ all correlating to the pair asymptote $5s+5p$. The paper closes with the conclusion.

\begin{figure}
    \includegraphics{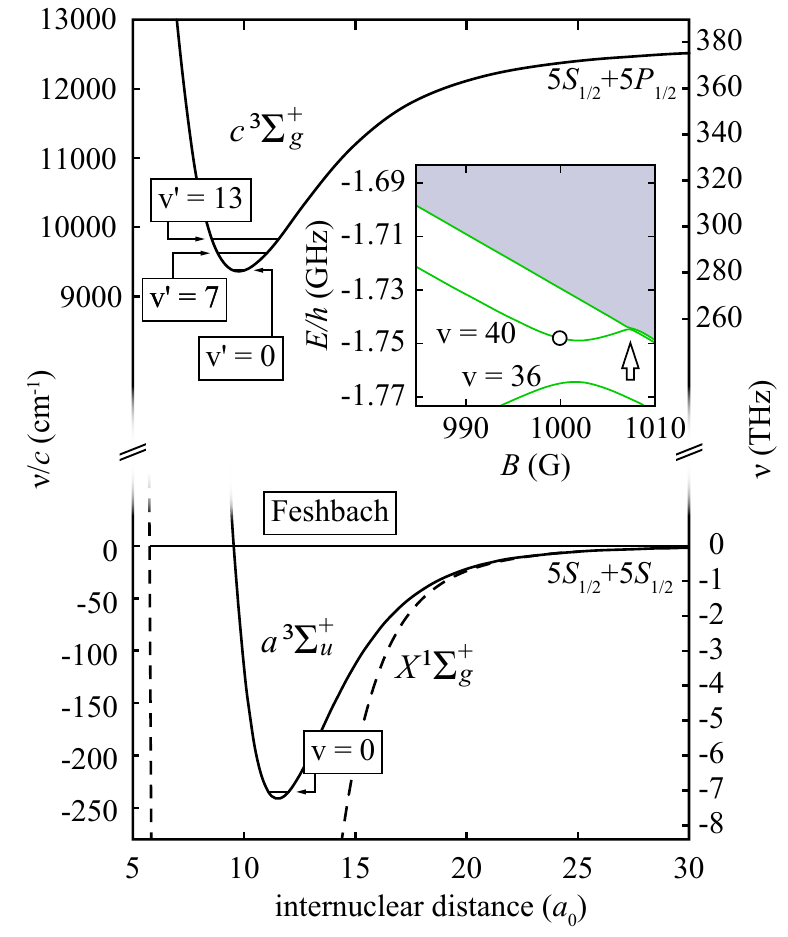}
	\caption{Molecular potentials \Xstate, \astate and \cstate and the relevant vibrational levels. The inset shows the level structure in the vicinity of the Feshbach resonance which is indicated by the arrow. The circle represents the frequency location of the Feshbach state at $B=999.9\,\text{G}$.}
	\label{fig:specScheme}
\end{figure}

\begin{figure}
	\includegraphics{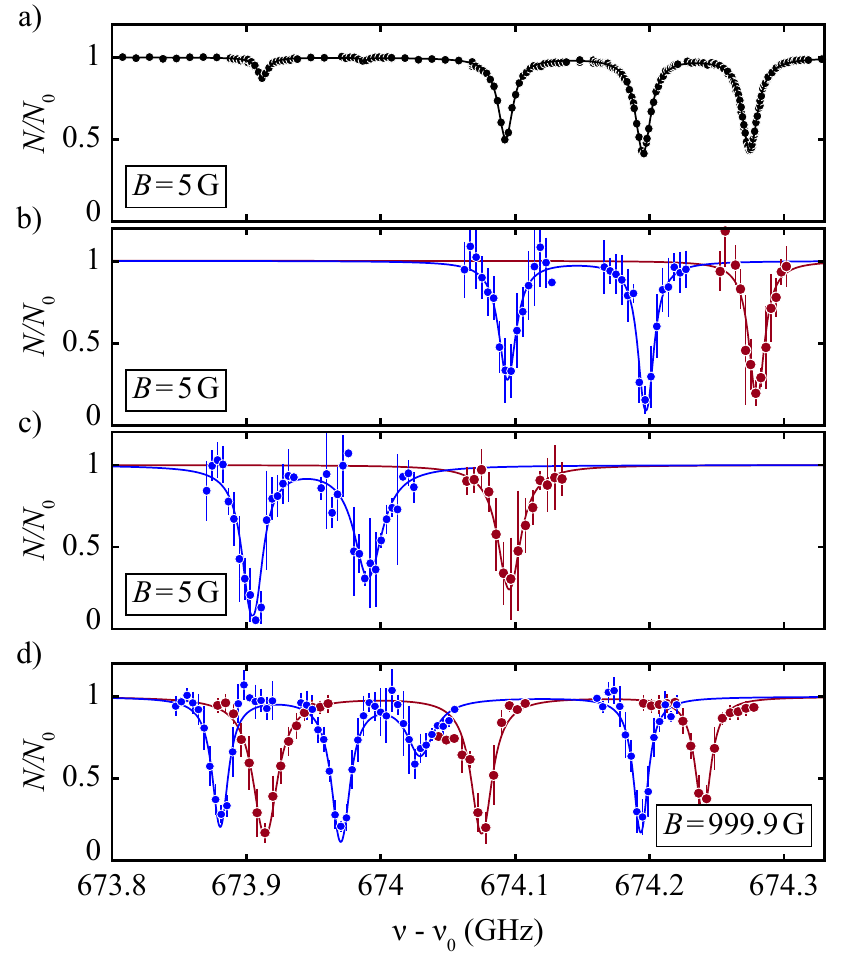}
	\caption{Spectra of the hyperfine levels of the \cstate, $0_g^-$, $\vib'=13$, $J'=2$ manifold. The data are taken for various initial levels of the \astate state: a) photoassociation of colliding atoms, b) $|a, R = 0 \rangle $, c) $|a,  R = 2 \rangle$ and d) Feshbach molecules (for the notation of the molecular states, see Tab.$\:$\ref{tab:groundstate}). The measurements a)-c) are carried out at a $B$-field of about $5\:\textrm{G}$ while d) is obtained at a $B$-field of $999.9\:\textrm{G}$. Blue circles correspond to $\pi$-transitions, red circles to $\sigma$-transitions, and black circles to a mixture of both. The bars give the standard mean error. For each resonance dip in b), c) and d) the laser power was adjusted individually to optimize the signal. The solid lines are model curve fits to our data. $\nu_0 = 294\,000\:\text{GHz}$.}
\label{fig:spectra}
\end{figure}

\section{Spectroscopic methods}
\label{sec:expSetup}

We start by introducing the two spectroscopic methods, photoassociation and photoexcitation, with which we investigate the substructure of the vibrational levels $v' = 0$, 7, 13 of the \cstate state (see Fig.~\ref{fig:specScheme}). This vibrational interval gives enough energy spread to test the modeling of hyperfine interaction.

\subsection{Short-range photoassociation spectroscopy}
\label{sec:photoassociation}

In order to carry out photoassociation spectroscopy we prepare a cold gas of about $2\times10^6$ $^{87}$Rb atoms in the electronic ground state at a temperature of about 1\,$\mu$K. The atoms are spin-polarized with total angular momentum $f_a=1$, $m_{fa}=-1$. They are held in a crossed optical dipole trap with a laser wavelength of $1064\,\textrm{nm}$ and trap frequencies of $\omega_{x,y,z}=2\pi\times(23,170,179)\,\textrm{Hz}$ in the three spatial directions, resulting in a density of about $3\times10^{13}\,\textrm{cm}^{-3}$. The atoms are exposed for a duration of $3\,\textrm{s}$ to the radiation of a grating-stabilized cw diode laser with a short-term linewidth of $\approx100\:\textrm{kHz}$. The laser is frequency-stabilized to a wavemeter (High Finesse WS7) and can be tuned over a range of $1003-1075\:\textrm{nm}$. The laser resonantly photoassociates colliding atom pairs to form \cstate molecules. This induces losses in the atom number which we measure via absorption imaging. The photoassociation beam is linearly polarized with an angle of $45^{\circ}$  with respect to the magnetic field axis and drives $\sigma^{\pm}$ and $\pi$-transitions with almost equal strength. At the location of the atomic sample the laser beam has a power of $\approx 30\,\text{mW}$ and a radius ($1/e^2$) of $0.28\,\text{mm}$. For details on the cold atom apparatus see Ref.~\cite{Schmid2012}. {Figure~\ref{fig:spectra}a)} shows a typical photoassociation signal at a magnetic field $B$ of about $5\,\textrm{G}$. As the frequency  of the photoassociation laser is scanned the remaining fraction of atoms $N/N_0$ is recorded. We note that throughout this publication, we do not report the measured transition frequencies but rather the term frequencies ($\nu$), i.e. the term energies divided by $h$. As reference for zero term frequency we choose the atomic dissociation limit at $0\,\text{G}$ for the atomic pair state $(f_a=1,m_{f_a}=+1)+(f_b=1,m_{f_b}=+1)$ of the $5S_{1/2} + 5S_{1/2}$ asymptote (see Fig.$\:$\ref{fig:specScheme}).

For the whole spectrum in {Fig.~\ref{fig:spectra}a)}, the laser intensity and pulse duration are kept constant. The data show resonance lines of hyperfine levels of the state \cstate , \zeroG, $\vib' = 13$,  $J' = 2$. Here $J'$ represents the quantum number of the total angular momentum  $\vec{J'}=\vec{L'}+\vec{S'}+\vec{R'}$, where $\vec{L'}$ is the total electronic orbital angular momentum, $\vec{S'}$ is the total electron spin and $\vec{R'}$ is the rotational angular momentum of the atom pair. We observe resonance linewidths of about $15\,\textrm{MHz}$, close to the natural linewidth of about $12\:\textrm{MHz}$, simply estimated as two times the width of the  atomic $5s-5p$ transition. The remarkable smoothness of the curve is due to small shot-to-shot fluctuations of the atom numbers of the prepared cloud of less than $\pm2\%$. In addition, the signals are surprisingly strong considering that we are performing short-range photoassociation where  Franck-Condon factors are generally not very favorable. The hyperfine splitting of the lines will be investigated in more detail in the sections \ref{sec:groundstate} and \ref{sec:modelCalc}.

\subsection{Photoexcitation spectroscopy}

\begin{figure}[htb]
    \includegraphics[]{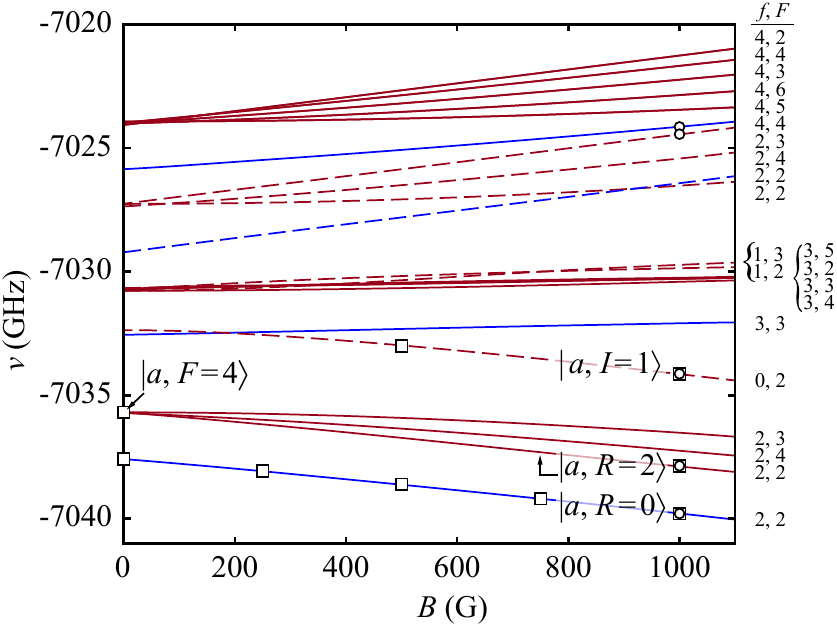}
    \caption{ Level structure of the \astate, $\vib=0$, $m_F = 2$ manifold. Lines are calculations based on a coupled-channel model \cite{Strauss2010}. Blue lines are states without rotational excitation ($R=0$), red lines are $R=2$ states. Dashed (solid) lines correspond to total nuclear spin $I = 1$ (3), respectively. The total angular momenta $f$ and $F$ are given for each level for $B=0\:\textrm{G}$. The circles at $B=999.9\:\textrm{G}$ represent dark-state spectroscopy measurements (see table \ref{darkstate} of the Supplemental Material for the precise frequency values), while the squares show level positions which are used as initial states for spectroscopy of the \cstate manifold.}
    \label{fig:groundstate}
\end{figure}

For photoexcitation spectroscopy we prepare a pure cloud of ultracold Rb$_2$ molecules such that all molecules are in the same pre-defined quantum state within the \astate potential. The molecules are irradiated for a duration of a few milliseconds by a cw grating-stabilized diode laser which resonantly excites them to \cstate levels, leading to molecular loss. We measure the remaining number of molecules by first dissociating them into ultracold atom pairs and then measuring the corresponding atom number via absorption imaging (for details see \cite{Takekoshi2011, Deiss2015}). By scanning the laser frequency from shot to shot, a resonance spectrum is recorded [see Fig.$\:$\ref{fig:spectra}b) to d)]. The photoexcitation laser has a short-term linewidth of $\approx100\,\text{kHz}$. We typically use a rectangular light pulse with a power of up to a few hundred $\mu$W at the location of the molecular sample, where the beam waist is about $1.1\,\text{mm}$. The light propagates orthogonally to the direction of the magnetic field which sets the quantization axis. By means of a half-wave plate we can choose the linearly polarized laser light to drive either $\pi$- or $\sigma^\pm$-transitions. The molecules are held in a 3D optical lattice of laser wavelength $1064\,\textrm{nm}$ with at most one molecule per lattice site (see \cite{Thalhammer2006, Lang2008, Deiss2014} for details). We either work with Feshbach molecules or with molecules in the vibrational ground state ($\vib=0$) of the \astate state, cf. Fig.~\ref{fig:specScheme}. The level positions of all used initial states are precisely known (\cite{Strauss2010} and Tab.$\:$\ref{tab:groundstate}, see also Fig.$\:$\ref{fig:groundstate}). The cloud of Feshbach molecules typically consists of $3{\times}10^4$ particles and is produced from ultracold atoms ($f_a=1$, $m_{fa}=+1$) via a sweep over a magnetic Feshbach resonance at $1007.4\,\text{G}$ \cite{Thalhammer2006, Lang2008, Deiss2014}. We ramp the $B$-field to $999.9\,\text{G}$ before photoexcitation of the Feshbach molecules.

The $\vib=0$ molecules are produced from the Feshbach dimers using stimulated Raman adiabatic passage (STIRAP), a coherent two-photon transfer process.  In our set-up we achieve a transfer efficiency of about 80\% (for details see \cite{Lang2008, Deiss2014}). By tuning the difference frequency of the STIRAP lasers, as well as by choosing an appropriate intermediate state we can precisely control which molecular level in the $\vib=0$ manifold is prepared. Within a few MHz we always find agreement with  measured  bound state energies listed in Ref.~\cite{Strauss2010} and with close-coupled channel calculations, respectively. Figure~\ref{fig:groundstate} shows such calculations as a function of the magnetic field.

\begin{table}
			\vspace{-6.5pt}
	    \caption{Quantum numbers and calculated term frequencies of various \astate, $\vib=0$ levels which we use as starting levels for the spectroscopy. The term positions $\nu$ are listed for a given magnetic field $B$. The middle column gives the acronyms for the levels.}
	    \label{tab:groundstate}
	\vspace{5.0pt}
    \centering
    \begin{tabular}{m{3mm} m{3mm} m{3mm} m{3mm} m{3mm} l c c }
    	\toprule
    	\toprule
    	$R$ & $I$ & $f$ & $F$ & $m_F$ & $\:\:$acronym$\:$              & $\nu $     & $B$   \\
    	    &     &     &     &       &                      & (GHz)      & (G)   \\ \midrule
    	0   & 3   & 2   & 2   & 2     & $\:\:$$|a, R = 0 \rangle$$\:$  & -7\,039.792 & 999.9 \\
    	2   & 3   & 2   & 2   & 2     & $\:\:$$|a, R = 2 \rangle$$\:$  & -7\,037.868 & 999.9 \\
    	2   & 1   & 0   & 2   & 2     & $\:\:$$|a, I = 1 \rangle$$\:$  & -7\,034.159 & 999.9 \\
    	2   & 3   & 2   & 4   & 4     & $\:\:$$|a, F =  4 \rangle$$\:$ & -7\,035.759 & $\sim 5$     \\
   	0   & 3   & 2   & 2   & 2     & $\:\:$$|a, F =  2 \rangle$$\:$ & -7\,037.596 & $\sim 5 $    \\ \bottomrule
   	\bottomrule
    \end{tabular}
\end{table}

Specifically, for our experiments we work with four different \astate , $\vib=0$ levels which, at 0$\:$G, are characterized by the quantum numbers given in Tab.$\:$\ref{tab:groundstate}. The central column 'acronym' of the table gives a short-hand notation for these four levels. $R$ denotes  molecular rotation, $I$ is the total nuclear spin, $f$  is the total angular momentum without rotation (i.e. $\vec{f} = \vec{I} + \vec{S} + \vec{L}$), and $F$ is the total angular momentum ($\vec{F}=\vec{f}+\vec{R}$). The magnetic quantum number is in general $m_F = 2$, except for $|a,F=4 \rangle$. During the preparation of this state, i.e. when the $B$-field is ramped down from $999.9$ to $\sim5\:\textrm{G}$ an optical Raman transition flips the $m_F$ quantum number from 2 to 4. The Raman transition is driven by the laser beams used to generate the three standing light waves for the cubic 3D optical lattice. These three standing light waves have mutually orthogonal, linear polarizations and pairwise relative frequency detunings of $220\:\textrm{MHz}$, $190\:\textrm{MHz}$ and $30\:\textrm{MHz}$, respectively.

Typical recordings for different initial states are shown in Fig.~\ref{fig:spectra}b)-d), where the fraction of remaining particles $N/N_0$ is plotted versus the frequency $\nu$ of the spectroscopy laser. When starting from molecules the resonance lines are generally measured individually, i.e. for each line the intensity and pulse duration of the spectroscopy laser light are adjusted such that a good signal-to-noise ratio is reached. Consequently, within a single spectrum, different laser intensities and exposure times are used.

In order to simulate the spectra we take into account that the lineshape of a photoexcitation line is determined by the exponential loss of the molecules
\begin{eqnarray}
    N(\nu) = N_0 \cdot \exp{\left(-\sum\limits_\textrm{i} \gamma(\nu,\nu_\mathrm{0,i})~t\right)},
    \label{eq:specLines}
\end{eqnarray}
with $N$ and $N_0$ being the remaining and initial particle numbers, respectively. $t$ is the time of laser exposure and $\gamma(\nu,\nu_{0,\textrm{i}})$ is the loss rate for a given laser frequency $\nu$ and excited level $i$ with resonance frequency $\nu_{0,i}$. Since in our experiment we work in a regime, where the transition lines are in general not saturated, the loss rate is expressed by the Lorentzian
\begin{eqnarray}
    \gamma(\nu,\nu_{0,i}) \propto \frac{\tilde{\Omega}^2/\Gamma}{1 + 4((\nu-\nu_\mathrm{0,i})/\Gamma)^2}.
    \label{eq:phScattering}
\end{eqnarray}

The natural decay rate of the excited state $\Gamma$ is expected to be approximately twice the rate of single rubidium atoms, $\Gamma \approx 2\times (2\pi \times 6\:\text{MHz})$. The Rabi frequency $\tilde{\Omega}$ is used as a free parameter. As shown in Fig.~\ref{fig:spectra} the measurements are well described by the corresponding lineshape fits. We use such fits only to determine the resonance frequencies $\nu_\mathrm{0,i}$.

The measurements of Fig.~\ref{fig:spectra}a)-c) investigate the same hyperfine levels of the \cstate, $\vib' = 13$, \zeroG, $J' = 2$ state at a $B$-field of a few gauss. For different initial states different excited levels can be observed, due to  selection rules. By choosing the polarization of the laser, $\pi$- or $\sigma$-transitions are driven. Within the spectroscopic resolution of a few $\textrm{MHz}$ in our experiments the transition lines towards the same excited levels are found on top of each other. This is an important consistency check for the spectroscopy and also shows that the term energies of the initial molecular states are precisely known.

Figure~\ref{fig:spectra}d) shows the spectrum in the same term-energy range at a $B$-field of $999.9\,\textrm{G}$ observed by photoexcitation of Feshbach molecules. Clearly, the Zeeman effect has split and shifted  the lines as compared to the case at low $B$-field. In fact, some of the lines are shifted outside the shown frequency window. Thus Fig.$\:$\ref{fig:spectra}d does not cover all observable $J'=2$ levels.

\begin{figure*}[htb]
	\centering
	\includegraphics{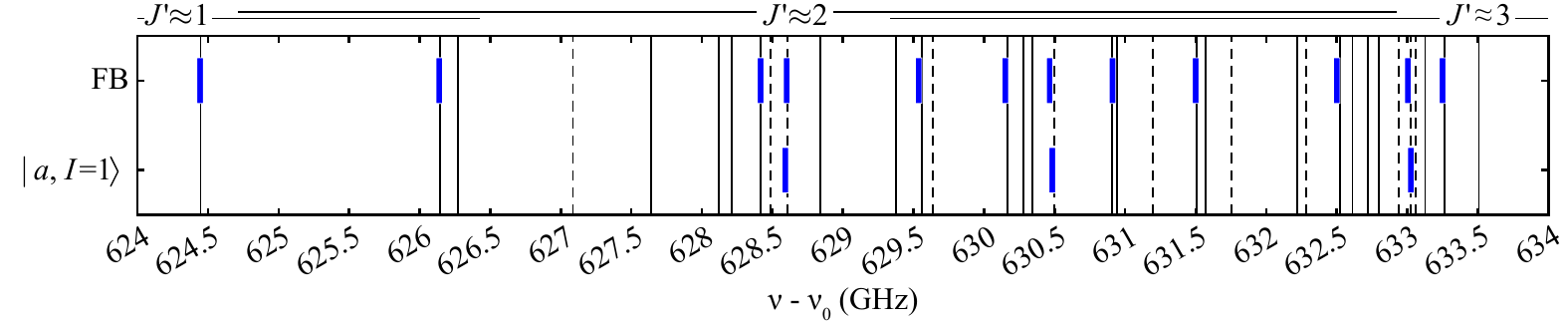}
	\caption{Level spectrum for \cstate, \oneG, $\vib'=13$, $m'_F = 2$ at $B=999.9\,\text{G}$. The measurements (blue long bars) are taken with $\pi$-polarized light. The initial states for the spectroscopy are indicated on the left. The vertical solid lines show calculations of the levels with $I'=3$, dashed ones with $I'=1$. Above the figure the ranges of the rotational states $J'$ are indicated. $\nu_0 = 294\,000\,\text{GHz}$.}
	\label{fig:v13_1000G}
\end{figure*}

\subsection{Spectroscopic calibration and uncertainty}

In our experiments we use two High Finesse WS7 wavemeters to measure the laser frequencies. In intervals of minutes these wavemeters are repeatedly calibrated to an atomic $^{87}$Rb reference signal at a wavelength of $780\,\text{nm}$. Both instruments have a specified absolute accuracy of $60\,\text{MHz}$. However, for difference frequency determinations within several $100\,\text{MHz}$ the accuracy is on the MHz level. Furthermore, we checked that over a period of several months the frequency readings for molecular transitions are reproducible for each wavemeter within $\pm10\,\text{MHz}$. Nevertheless, despite the calibration with Rb, we observed a relative frequency offset between the two wavemeters of $48\,\text{MHz}$ when measuring wavelengths at around 1040$\,\textrm{nm}$, i.e. in the range relevant for the present work. Therefore, in order to be able to work with the measured frequencies without ambiguity we have arbitrarily chosen one of the two wavemeters to be the reference and thus correct the reading of the other wavemeter correspondingly.

Our spectroscopy lasers are frequency-stabilized to the wavemeters, with an update rate of about $10\,\text{Hz}$. In total, we obtain a frequency stability of $\pm4\,\text{MHz}$, mainly determined by shot-to-shot readout fluctuations of the wavemeters and a smaller contribution of laser frequency drifts between updates.

As noted earlier in section \ref{sec:photoassociation}, the term frequencies ($\nu$) that we report in this publication are referenced with respect to the atomic dissociation limit at $0\,\text{G}$ for the atomic pair state $(f_a=1,m_{f_a}=+1)+(f_b=1,m_{f_b}=+1)$. This reference is $8.543\,\text{GHz}$ below the $5S_{1/2} + 5S_{1/2}$ threshold for which hyperfine interaction is ignored. For completeness, we note that the Feshbach state at a magnetic field of $999.9\,\text{G}$ is located at $-1.748\,\text{GHz}$ (see inset of Fig.$\:$\ref{fig:specScheme}). The lowest level in the \astate vibrational ground state ($R = 0$, $F = 2$, $m_F = 2$) at $B=0\,\textrm{G}$ has a frequency of $-7\,037.587\,\text{GHz}$.

\section{Spectroscopic data}
\label{sec:groundstate}

\subsection{Spectroscopy at 999.9$\:$G}

We start our investigation by revisiting the $\vib'=13$ vibrational level of the \cstate state, which already has been studied in our previous work by Takekoshi~\emph{et al.}~\cite{Takekoshi2011}. Since in Ref.$\:$\cite{Takekoshi2011} an inconsistency between the experiment and the theoretical model was noted we repeat and extend the measurements in order to answer the remaining questions. Figure~\ref{fig:v13_1000G} (upper spectrum) shows the new data for the \oneG state obtained from one-photon spectroscopy with $\pi$-polarized light, starting with Feshbach (FB) molecules at a magnetic field of $B=999.9\,\text{G}$. The relative frequency positions of the lines are determined with an improved uncertainty of about $5\,\textrm{MHz}$. All observed resonance positions are listed in Tab.$\:$\ref{tab:v13a} of the Supplemental Material (see also tables~\ref{tab:v0} and \ref{tab:v7} for the results on $\vib'=0$ and $\vib'=7$, respectively). Apart from a global frequency shift of about $80\,\textrm{MHz}$ (due to the fact that a different wavemeter was used) the observed spectrum of Fig.$\:$\ref{fig:v13_1000G} is very similar to the one of our previous work \cite{Takekoshi2011}. Compared to the analysis in \cite{Takekoshi2011} we change the assignment of several lines. These lines (in particular the one at about $294.6305\,\text{THz}$ which could not be explained in Ref.$\:$\cite{Takekoshi2011}) do not correspond to a total nuclear spin $I'=3$ but rather to $I'=1$. In order to show this, we repeat the spectroscopic measurements for these excited levels, however, starting now from a  $I = 1$ level. Specifically, this level belongs to the vibrational ground state of the \astate potential and its properties are known from accurate close-coupled channel calculations \cite{Strauss2010}. It has the quantum numbers $|I = 1,\:f = 0,\:R = F = 2,\:m_F = 2 \rangle \equiv |a, I = 1 \rangle$ and its position is marked in Fig. \ref{fig:groundstate}. It is experimentally prepared via STIRAP using the \cstate, $\vib'=13$, $I'=1$ level at $\nu = 294\,630.484\,\text{GHz}$ as intermediate level. This, as a byproduct demonstrates that \astate, $\vib=0$ molecular states with different nuclear spins can be prepared via STIRAP by choosing appropriate intermediate states (The preparation of $I=3$ levels has been shown e.g. in \cite{Lang2008, Deiss2014}.).

We now use $|a, I = 1 \rangle$ as a starting level to perform spectroscopy on the $\vib'=13$ manifold. For this we scan over parts of the frequency range shown in Fig.~\ref{fig:v13_1000G} and observe three $I'=1$ levels but no $I'=3$ levels [see Fig.~\ref{fig:v13_1000G} (lower spectrum)]. These three levels have also been observed in the scan with Feshbach molecules. The fact that we can observe both $I'=1$ and $I'=3$ levels starting from Feshbach molecules indicates that the Feshbach state is a mixture of nuclear spins $I=1$, $3$. Indeed, this is confirmed by our close-coupled channel calculations~\cite{Strauss2010} which yield that the Feshbach state at $999.9\:\textrm{G}$ has a \astate component of 84\% and a \Xstate component of 16\%. The \astate component is a mix of $I = 1$, 3 while the \Xstate component of the Feshbach molecules has $I = 2$. In total, the contribution of $I=1$ to the Feshbach state is about $22\%$, which is sizable. The thin vertical lines in Fig.~\ref{fig:v13_1000G} are calculations for all levels of \cstate, $\vib'=13$, $\oneG$ with $m'_F = 2$ in the given frequency range, based on our model which will be introduced in section~\ref{sec:modelCalc}. Clearly, the observed lines are a subset of the calculated levels. We have verified that we detect all predicted levels for which a $\pi$-transition is allowed by the selection rules. Those levels in Fig.~\ref{fig:v13_1000G} which are not observed are only visible when using $\sigma$-polarized light, or they correspond to energy levels with total angular momentum quantum number $F' > 4$. These levels cannot be addressed following the selection rule $\Delta F = 0$, $\pm 1$ (which, however, only strictly holds at $B=0$). Around the measured resonance frequency of $294.633012\,\text{THz}$ two $I'=1$ energy levels are actually predicted, but they cannot be resolved. To summarize, we now find consistent agreement between theory and experiment.

\subsection{Spectroscopy at low magnetic fields}

Next, we carry out spectroscopy of the \cstate, $\vib'=13$, \oneG manifold at low magnetic fields $B$ of just a few G. In our previous work of Ref.$\:$\cite{Takekoshi2011}, such measurements were difficult and only produced poor spectroscopic signals with large uncertainties. These measurements were carried out with Feshbach molecules for which the ramp-down of the $B$-field is passing over a number of avoided level crossings. This led to instabilities and low particle numbers. In the present work, we overcome this problem by either using short range photoassociation spectroscopy or by working with Rb$_2$ molecules in the vibrational ground state for which avoided level crossings are rare cases (see Fig.$\:$\ref{fig:groundstate}).

\begin{figure*}[htb]
	\centering
	\includegraphics{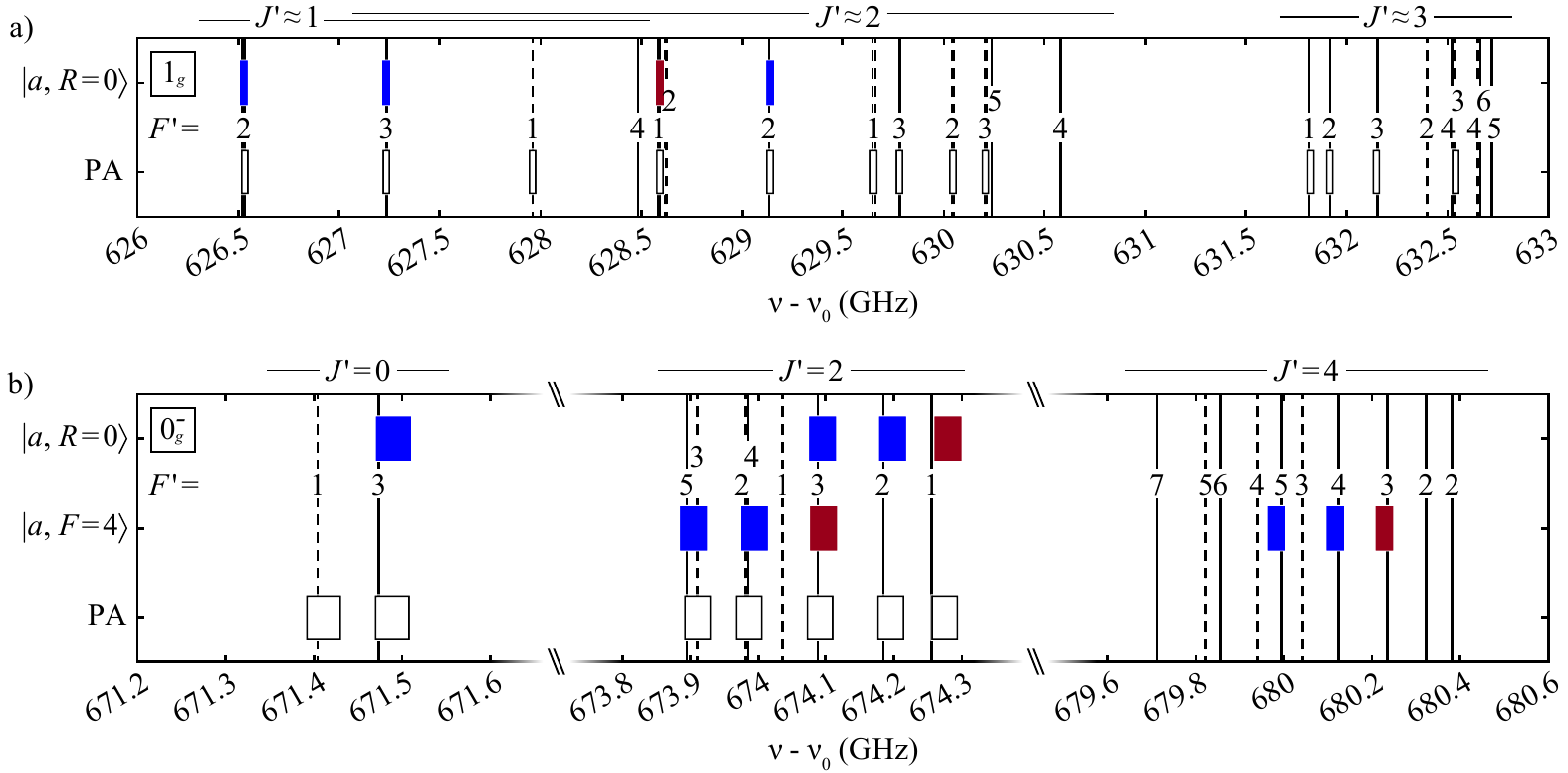}
	\caption{Spectra for \cstate, $\vib'=13$, \oneG (a) and $0_g^-$ (b) at $B \approx 5\,\text{G}$. The boxes represent measurements while calculations are represented by lines. The width of the boxes is always $40\,\text{MHz}$ and helps to identify the different frequency scales in the figure. Blue (red)  boxes correspond to measurements with $\pi$ ($\sigma$)-polarized light. White boxes illustrate a mixture of both. The initial states are given on the left. Solid lines belong to $I'=3$, dashed ones to $I'=1$. Above the figure the ranges of the rotational states $J'$ are indicated. $\nu_0 = 294\,000\,\text{GHz}$.}
	\label{fig:v13_0G}
\end{figure*}

The method of photoassociation (PA) spectroscopy has already been discussed in section~\ref{sec:photoassociation}. Figure~\ref{fig:v13_0G}a (lower spectrum) shows the derived term frequencies from the observed transitions together with calculated level positions. The spectroscopy is carried out with linearly polarized light which can equally drive $\pi$- and $\sigma$-transitions. Again, the agreement between the data and the model calculations is quite good since all observed levels can be assigned. Unobserved levels are not accessible due to selection rules, such as $\Delta F = 0$, $\pm1$ which holds strictly because of the low $B$-field.

In order to check for consistency, we now apply the second method, where we start at $B=999.9\:\text{G}$ with Rb$_2$ molecules in the vibrational ground state either in level $|a, R = 0\rangle$ or $|a, R = 2\rangle$. The magnetic field is ramped down to the desired value and one-photon spectroscopy towards \cstate is performed. Since avoided crossings are absent in  Fig.$\:$\ref{fig:groundstate} one might expect a fully-adiabatic transfer. However, for the particular case of $|a, R = 2\rangle$ we find that a two-photon induced spin flip which is driven by the lattice lasers changes the state  $|a, R = 2\rangle$ to  $|a, F = 4\rangle$ during the ramp down, see also \cite{Deiss2014}. The term frequencies of the initially prepared states $|a, R = 0\rangle$ and  $|a, F = 4\rangle$ for the magnetic field of $5\:\textrm{G}$ are determined from close-coupled channel calculations for the \astate and \Xstate states \cite{Strauss2010}, see Tab.~\ref{tab:groundstate}.

Using the initial state $|a, R = 0\rangle$ and starting from low laser frequencies we record a spectrum for $\vib' = 13$, \oneG until four transition lines are observed. The four corresponding levels are shown in Fig.~\ref{fig:v13_0G}a (upper spectrum). We use either $\pi$- or $\sigma$-polarized light, as indicated by the blue and red plot symbols. Indeed, according to the selection rules the four transition lines correspond to the energetically lowest four accessible levels of $\vib' = 13$. The resonance positions nicely confirm both the results obtained from the photoassociation of atoms and the calculated binding energies of the levels at $B = 5\:\textrm{G}$.

Next we measure the spectra for the $\vib'=13$, \zeroG manifold, using photoassociation and photoexcitation. For photoexcitation we start either from the state $|a, R = 0\rangle$ or from the state $|a, F = 4\rangle$. The spectra (see Fig.$\:$\ref{fig:v13_0G}b)  are consistent with each other and also consistent with the calculations. For \zeroG the $J'$ quantum number is quite good and therefore both the photoassociation and the photoexcitation from $|a, R = 0\rangle$ can only reach $J' = 0$, 2, because the starting levels have $J \approx 1$ and their total parity is even. By contrast, from $|a, F = 4 \rangle$ we can reach  $J' = 4$. We have verified that the transition lines indicated in red in Fig.$\:$\ref{fig:v13_0G}b can only be observed with $\sigma$-polarized light.

Additionally to $\vib'=13$, we have also measured the spectra for $\vib'=0$ and 7 which look alike and give similar results. All derived term frequencies are provided in tables~\ref{tab:v0} to \ref{tab:v13a} of the Supplemental Material.

\subsection{Spectroscopy at intermediate magnetic fields}

\begin{figure}[htb]
    \includegraphics[width=\columnwidth] {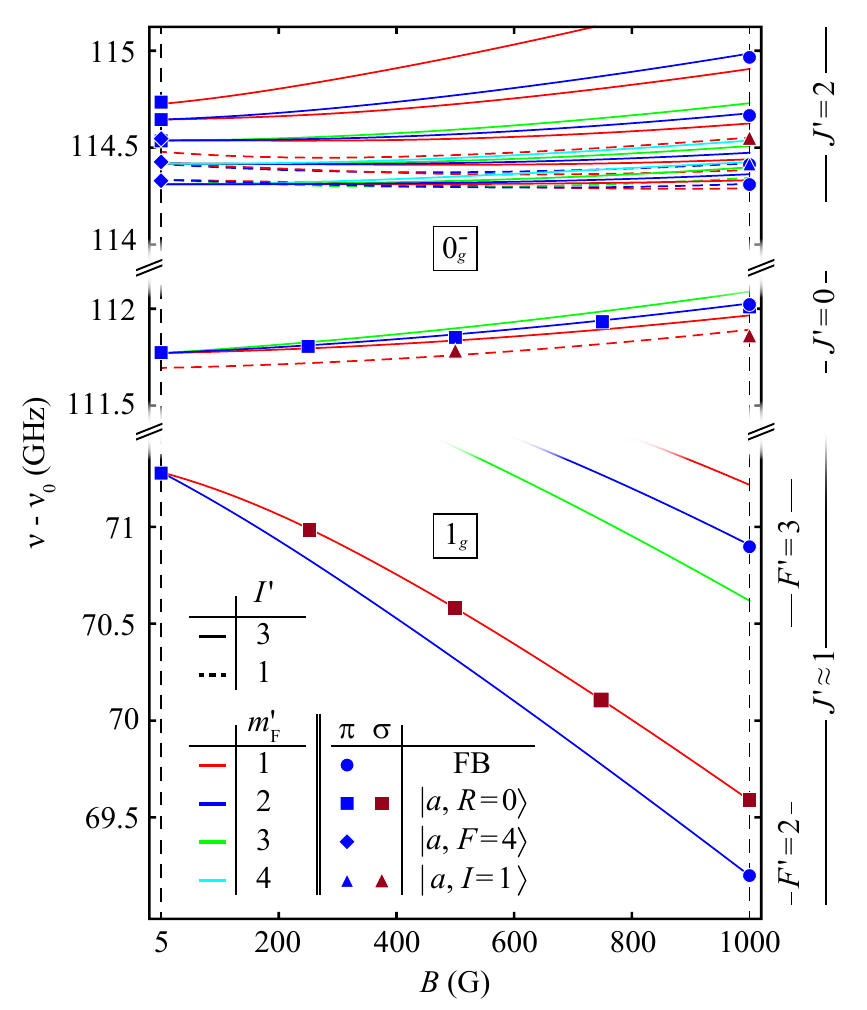}
    \caption{Zeeman shifts of \cstate, $\vib'=0$ energy levels. Symbols are measurements while lines are calculations. The legend indicates for each plot symbol the initial state and the polarization corresponding to the observed transition. Dashed lines indicate levels with nuclear spin $I'=1$ while the solid lines represent $I'=3$. Furthermore, the $m'_F$ quantum numbers of the molecular levels are color-coded as given in the legend. On the right, groups of levels are assigned with the corresponding $F'$ ($J'$) quantum numbers for a magnetic field of  $\approx0\:\textrm{G}$ ($\approx1000\:\textrm{G}$). Dashed vertical lines mark the magnetic field values of $5\:\textrm{G}$ and $999.9\:\textrm{G}$. $\nu_0 = 281\,000\,\text{GHz}$.}
    \label{fig:v0Transition}
\end{figure}

For some selected levels within the \cstate potential we studied the Zeeman shift in more detail, i.e. besides the term energies at $B =999.9$\,G and at $B \approx 5$\,G  we also carried out measurements for various magnetic field strengths in between. For this, we first produce molecules at $B=999.9\:\textrm{G}$. Subsequently, the magnetic field is lowered to the desired value. Figure~\ref{fig:v0Transition} shows the results for the case of $\vib'=0$. Three levels were investigated: one level of state \oneG with quantum numbers $I'=3$, $J' \approx 1$, $F'=2$ and two levels of state \zeroG with $J' = 0$ and $I'=3, 1$, respectively.

For the $I'=1$ state no measurements below $500\:\textrm{G}$ could be carried out because at such magnetic fields strong particle losses occurred. Possibly, these losses are due to photoexcitation of the molecules by the optical lattice lasers.

The data of Fig.~\ref{fig:v0Transition} are obtained by adding the term energies of the initially prepared states to the measured transition energies. The term energies of the initially prepared states for magnetic fields below $999.9\:\textrm{G}$ are calculated using close-coupled channel calculations for the \astate and \Xstate states \cite{Strauss2010}. The levels and magnetic fields that were used for the spectroscopy are marked with square plot symbols in Fig.~\ref{fig:groundstate}. Figure~\ref{fig:v0Transition} shows that all experimentally determined Zeeman shifts of the \cstate, $\vib'=0$ level are well described by our effective Hamiltonian model of the following section \ref{sec:modelCalc}.

\section{Model calculations}
\label{sec:modelCalc}

\subsection{Effective Hamiltonian}
In order to interpret and analyze the measured \cstate spectra we use a simple molecule model which has been described in more detail in Ref.~\cite{Denschlag2016} (see also \cite{Takekoshi2011}). Within the model the molecule is treated as a rigid rotor where two separate neutral atoms are held at a fixed distance and rotate about their common center-of-mass. Each atom has a well defined electronic orbital momentum, i.e. for the \cstate state one has an $s$-orbital and the other one has a $p$-orbital. Given a vibrational level $\vib'$ of the real molecule, the model parameters are adjusted such that the model describes well the level structure of the \cstate molecule within the particular manifold of that vibrational level. Specifically, we obtain three sets of model parameters for the vibrational quantum numbers  $\vib'=0$, 7 and 13 studied.
The Hamiltonian is given by
\begin{equation}
    H = E^{\vib'}_0 + H_\mathrm{ss} + H_\mathrm{rot} + H_\mathrm{hf} + H_\mathrm{Z}.
    \label{eqn:Hamiltonian}
\end{equation}
It acts on a Hilbert subspace where the total electronic spin is $S = 1$ and the projection of the total orbital angular momentum $\vec{L}$ onto the internuclear axis is $\Lambda = 0$ which corresponds to a $\Sigma$ state. The first term, $E^{\vib'}_0 $, of the Hamiltonian $H$ is the energy offset for the vibrational level $\vib'$. The term $H_\mathrm{ss} = -2\lambda_{\text{v}'}(\vec{n} \cdot \vec{S})^2/\hbar^2$ is an effective spin-spin interaction that gives rise to the splitting of the \zeroG and \oneG components. Here, $\vec{n}$ is the unit vector along the internuclear axis of the diatomic molecule, $\vec{S} $ is the total electronic spin vector and $\lambda_{\text{v}'}$ denotes the effective spin-spin interaction parameter for the given vibrational level $\text{v}'$. The term $H_\mathrm{rot} = B_{\text{v}'} \vec{R}^2/\hbar^2$ represents the rotational energy of the molecule. $\vec{R}$ is the rotational angular momentum and $B_{\vib'}$ is the rotational constant. The hyperfine interaction, i.e. the coupling between the total nuclear spin $\vec{I}$ and the total electronic spin $\vec{S}$ is described by $H_\mathrm{hf} = (b_F - \frac{1}{3}c) \vec{I} \cdot \vec{S}/\hbar^2 + c (\vec{I}\cdot\vec{n})(\vec{S}\cdot\vec{n})/\hbar^2$, where $b_F$ is the Fermi contact parameter and $c$ is the anisotropic hyperfine parameter. Following Ref.~\cite{Takekoshi2011} we split $H_\mathrm{hf} = H^\mathrm{diag}_\mathrm{hf} + H^\mathrm{off}_\mathrm{hf}$ into a diagonal and an off-diagonal term with respect to both operators $\vec{I}\cdot\vec{n}$ and $\vec{S}\cdot\vec{n}$. Specifically, we define $H^\mathrm{diag}_\mathrm{hf} = c^\mathrm{diag} (\vec{I}\cdot\vec{n}) (\vec{S}\cdot\vec{n}) /\hbar^2$ with $c^\mathrm{diag}=b_F+\frac{2}{3}c$ and $H^\mathrm{off}_\mathrm{hf} = c^\mathrm{off} (\vec{I}\cdot\vec{S} - (\vec{I}\cdot\vec{n}) (\vec{S}\cdot\vec{n})) /\hbar^2$ with $c^\mathrm{off}=b_F-\frac{1}{3}c$. Finally, the last term $H_\mathrm{Z} = - \mu_B \ (g_S \ \vec{S} + g_L \ \vec{L} ) \cdot \vec{B}/\hbar$ is the Zeeman interaction with the magnetic field $\vec{B}$. Here, $\mu_B$ is the Bohr magneton and $g_S = 2$, $g_L = 1$ are the $g$-factors of the electronic spin and orbital angular momentum, respectively. We neglect Zeeman interaction of the nuclear spins or of molecular rotation. Furthermore we omit spin-rotation interaction, since it will be very small for the low rotational angular momenta involved in this work.

\begin{figure}
    \includegraphics[width=\columnwidth]{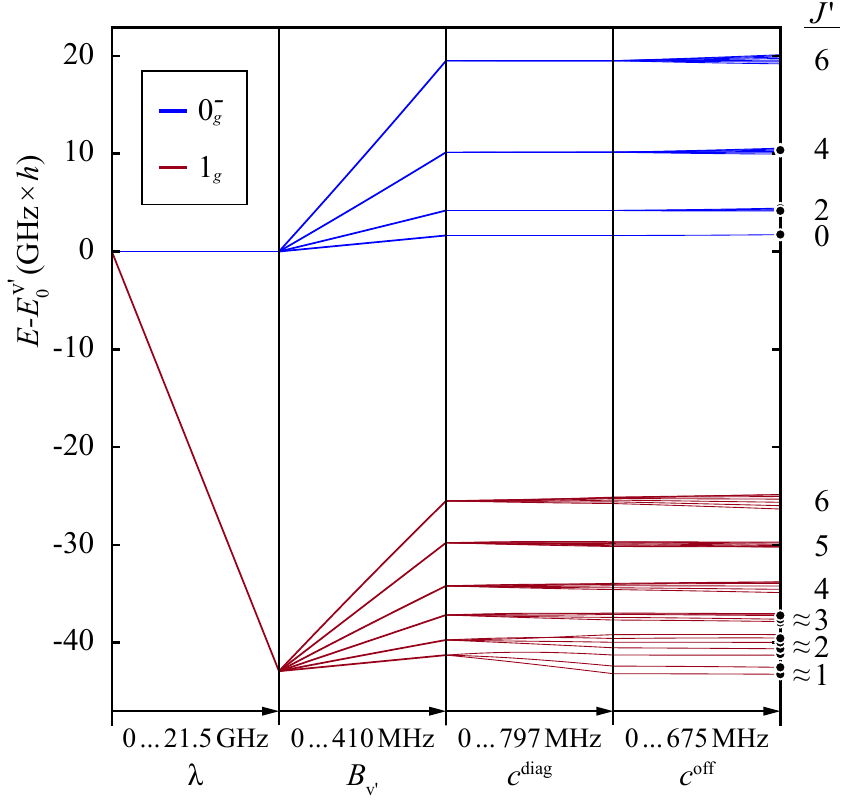}
    \caption{Influence of various interaction terms on the level spectrum. Shown are calculations for $\vib'=13$, $I'=3$,  $m'_F = 2$ at a magnetic field of $B\approx5\:\textrm{G}$. From left to right the effective spin-spin coupling, the rotation, the diagonal and the off-diagonal contributions of the hyperfine interaction are subsequently turned on to their values given in table~\ref{tab:results}. The circles on the right represent experimental results.}
    \label{fig:showParameters}
\end{figure}

Figure~\ref{fig:showParameters} illustrates the effect of the interactions $H_\mathrm{ss}$, $H_\mathrm{rot}$ and $H_\mathrm{hf}$ on the level structure. Their coupling constants $\lambda_{\text{v}'}$, $B_{\vib'}$, $c^\mathrm{diag}$ and $c^\mathrm{off}$ are subsequently turned on as we move from left to right in Fig.~\ref{fig:showParameters}. The effective spin-spin interaction $H_\mathrm{ss}$ leads to the large splitting of about $40\:\textrm{GHz}$ of the \zeroG and \oneG states that we observe in our experiments. Next, molecular rotation splits up levels with different $J'$ quantum numbers with $B_{\vib'} J'(J'+1)$, where $B_{\vib'} \approx 400\,\text{MHz}\times h$. Finally, the hyperfine interaction is added. We note that the diagonal part $c^\mathrm{diag} \approx 800\,\text{MHz}\times h$ has almost no influence on the $\zeroG$ levels of \cstate. In contrast, the $\zeroG$ levels generally split under the influence of the off-diagonal hyperfine interaction $c^\mathrm{off}$. We use these facts to extract the parameter $c^\mathrm{off}$ from our measured spectra with low uncertainty. Similarly, the  $J'=1$ level of \oneG is susceptible for $c^\mathrm{diag}$ but hardly for $c^\mathrm{off}$. Figure \ref{fig:showParameters} also indicates that for state \oneG~the rotational energy at low $J'$ is comparable to the hyperfine energy. Thus $J'$ will be in this case no good quantum number, which can also be seen from the respective expectation values included in the tables \ref{tab:v0}-\ref{tab:v13a} of all experimental data in the Supplemental Material.

\subsection{Results for molecular parameters}
\label{ssec:results}

We determine the model parameters $E^{\vib'}_0$, $\lambda_{\text{v}'}$, $B_{\vib'}$, $c^\mathrm{diag}$ and $c^\mathrm{off}$ by diagonalizing the Hamiltonian of Eq.~(\ref{eqn:Hamiltonian}) and fitting the resulting energy spectrum to the measurements. For each of the three vibrational levels $\vib' = 0$, 7, and 13 we obtain a set of fit parameters given in Tab.~\ref{tab:results} with their respective uncertainties. In general, the overall agreement between the measurements and the model calculations is quite good as can be seen from the root mean square errors and by the comparisons of the hyperfine level energies given in tables \ref{tab:v0}-\ref{tab:v13a} of the Supplemental Material (see also Figs.$\:$\ref{fig:v13_1000G} and \ref{fig:v13_0G} and discussion in section \ref{sec:groundstate}).

As expected, the rotational constant $B_{\vib'}$ decreases with $\vib'$. In contrast to that, the effective spin-spin interaction parameter $\lambda_{\vib'}$ increases with $\vib'$, because the mainly perturbing state $^1\Pi_g$ is energetically above \cstate as will be discussed in section \ref{sec:ccModel}. This shifts the $1_g$ component of \cstate below the $0_g^-$ component and its influence increases for higher $\text{v}'$ coming closer to $^1\Pi_g$.

For the vibrational level $\vib'=13$ we can compare our results to the ones from \cite{Takekoshi2011}. The value for the parameter $2\lambda_{\vib'}$ reported in \cite{Takekoshi2011} is about $4\:\textrm{GHz}$ larger as compared to our results. No error limits were reported in \cite{Takekoshi2011}. We find good agreement with the rotational constant $B_{\vib'} = 412\:\textrm{MHz}\times h $ of \cite{Takekoshi2011} and fair agreement for $c^\mathrm{diag}$ which is reported as $832\:\text{MHz}\times h$ in \cite{Takekoshi2011}.  Our current value for $c^\mathrm{diag}$ is  $c^\mathrm{diag} = b_F + \frac{2}{3} c = 797\,\text{MHz}\times h$ (see table \ref{tab:results}). The parameter $c^\mathrm{off}$ (and therefore also $c$), however, could not be determined in \cite{Takekoshi2011} because the quality of the data, especially for the \zeroG spectrum, was not sufficient. In our current work we are able to extract $c^\mathrm{off}$ from our measurements and we can therefore report the anisotropic hyperfine parameter $c$ for the first time. Its value of $c = 122 \pm 16\:\textrm{MHz}\times h$  (for $\vib' = 13$) is much smaller than $b_F$ but still sizable. Judging from table~\ref{tab:results}, $c$ and $b_F$ do not seem to depend significantly on the vibrational quantum number.

\begin{table}[]
			\vspace{-6.5pt}
		\caption{Fit results for the investigated vibrational levels $\vib'=0$, 7, 13. In addition to the model parameters we present the root mean square error (rmse) which quantifies the average deviation of the measured  data (frequency) to the fit curve.  All units are in $\text{MHz}{\times}h$. }
		\label{tab:results}
	\vspace{5pt}
	\centering
	\begin{tabular}{l rl @{~~} rl @{~~} rl}
		\toprule
		\toprule
		               & \multicolumn{2}{c}{$\vib'=0$} & \multicolumn{2}{c}{$\vib'=7$} & \multicolumn{2}{c}{$\vib'=13$} \\ \midrule
		$B_{\vib'}$:   & 430           & $\pm$ 1       & 420           & $\pm$ 2       & 410           & $\pm$ 1        \\
		$b_{F}$:       & 714           & $\pm$ 6       & 726           & $\pm$ 13      & 715           & $\pm$ 9        \\
		$c$:           & 127           & $\pm$ 12      & 107           & $\pm$ 23      & 122           & $\pm$ 16       \\
		$\lambda_{\text{v}'}$:     & 19\,219       & $\pm$ 4       & 20\,361       & $\pm$ 8       & 21\,454       & $\pm$ 5        \\
		$E^{\vib'}_0$: & 281\,109\,954 & $\pm$ 12      & 288\,538\,549 & $\pm$ 19      & 294\,669\,749 & $\pm$ 9        \\
		rmse:          &   \multicolumn{2}{c}{11.4}    &   \multicolumn{2}{c}{4.3}    &    \multicolumn{2}{c}{13.9}    \\ \bottomrule
		\bottomrule
	\end{tabular}
\end{table}

We now compare our results for $c$ and $b_F$ to the recent theoretical work of Lysebo \emph{et al.} \cite{Lysebo2013}. For the relevant internuclear distances the value of $b_F$ was calculated to vary between $850$ and $810\,\text{MHz}\times h$, which agrees with the  rule-of-thumb estimate of
$b_F \approx A_{HF}/ 4$ \cite{Bai2011,Denschlag2016}, where $A_{HF} \approx  3.42\:\textrm{GHz} \times h$  is the atomic hyperfine constant for $^{87}$Rb in the electronic ground state \cite{Arimondo1977}. These theoretical predictions for $b_F$ are not too far from our values of $b_F \approx 720\,\text{MHz}\times h$ in table~\ref{tab:results}. However, $c$ is predicted to be less than $10\,\text{MHz}\times h$ in \cite{Lysebo2013}, which is in clear disagreement with our measurements. In general, the uncertainties of our fit parameters are quite low (see table \ref{tab:results}) which indicates that our simple  model is a good description for the molecular level structure within a vibrational level.

In order to check for consistency of the analysis we carry out fits individually for high ($999.9\:\text{G}$) and low ($5\:\text{G}$) magnetic fields, and for the combined data of both magnetic fields. Table \ref{tab:diffB} shows the obtained parameter sets for $\text{v}' = 13$. The root mean square error for the combined fit is significantly larger than for the separated fits. This is probably related to the fact that the derived hyperfine parameters are slightly outside the estimated error limits between the low and high field fits. A possible reason is that the Zeeman effect is not modeled sufficiently well by only applying the $g$-factor of a free electron spin. Extension for bound electrons with spin-orbit interaction or adding rotational and nuclear Zeeman effect might be appropriate. But for a final conclusion one would need more data on the Zeeman effect.

\begin{table}  
			\vspace{-6.5pt}
		\caption{Fit results for $\vib'=13$ at different magnetic fields. The first and second column show the separate fit results for $B = 5\,\text{G}$ and $B = 999.9\,\text{G}$. For the final column all data were fitted simultaneously (c.f. Tab.~\ref{tab:results}). Units are in $\text{MHz}\times h$.}
		\label{tab:diffB}
	\vspace{5pt}
    \begin{tabular}{l r l r l r l}
    	\toprule
    	\toprule
    	             & \multicolumn{2}{c}{$B=5\,$G} & \multicolumn{2}{c}{$B = 999.9\,$G} & \multicolumn{2}{c}{combined} \\ \midrule
    	$B_{\vib'}$: & 410.0   & $\pm$ 0.4          & 410.8   & $\pm$ 0.7                & 410.3   & $\pm$ 0.8          \\
    	$b_F$:       & 726     & $\pm$ 5            & 712     & $\pm$ 8                  & 715     & $\pm$ 9            \\
    	$c$:         & 106     & $\pm$ 9            & 122     & $\pm$ 12                 & 122     & $\pm$ 16           \\
    	$\lambda_{\text{v}'}$:   & 21\,454 & $\pm$ 2            & 21\,450 & $\pm$ 5                  & 21\,454 & $\pm$ 5            \\
    	rmse:        &   \multicolumn{2}{c}{6.0}    &      \multicolumn{2}{c}{8.9}       &   \multicolumn{2}{c}{13.9}   \\ \bottomrule
    	\bottomrule
    \end{tabular}
\end{table}

\begin{figure*}
    \includegraphics{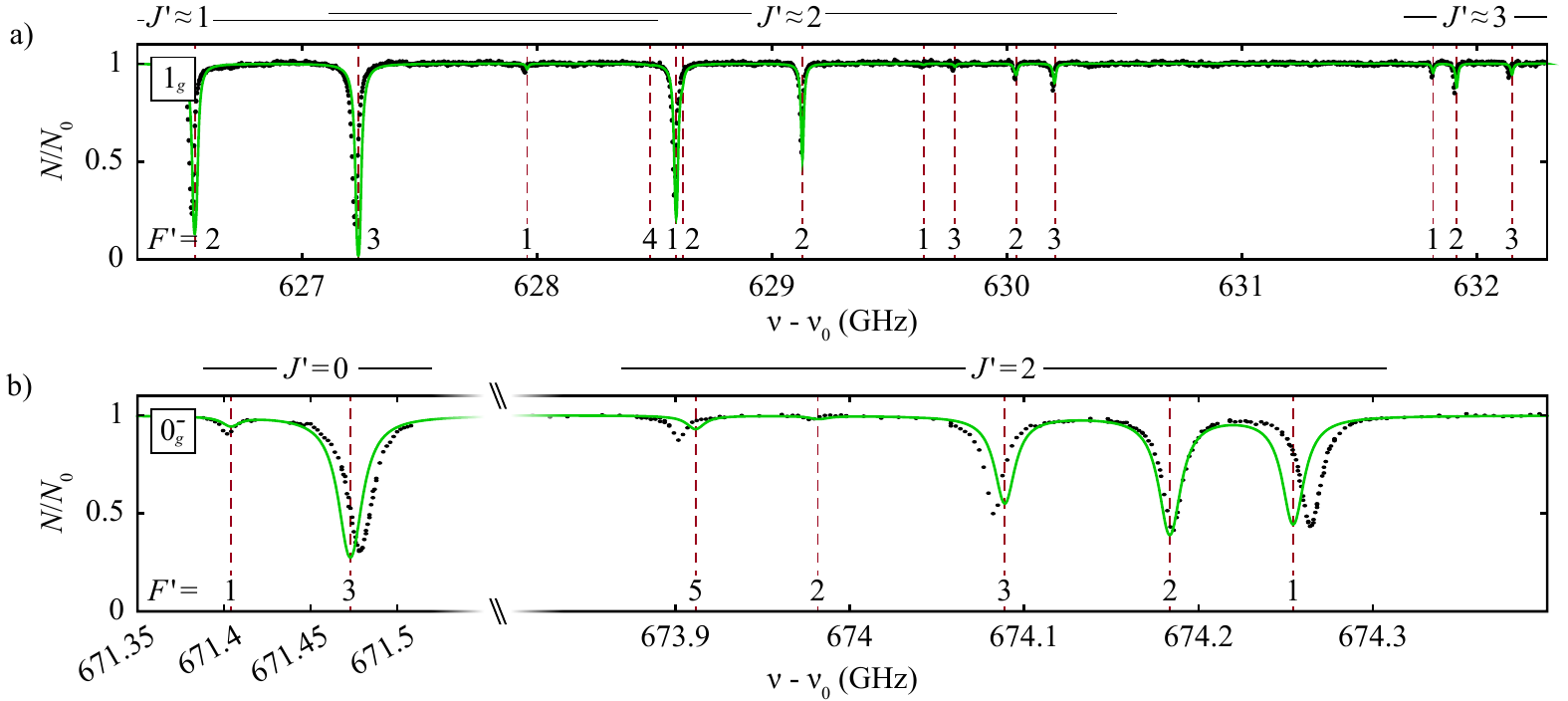}
    \caption{Relative transition strengths of lines. Shown are experimental loss spectra (dots) by photoassociation towards the states \cstate, $\vib'=13$ \oneG (a) and \zeroG (b) for fixed laser power and exposure time at $B\approx5\,\text{G}$. The solid lines are calculations. The vertical dashed lines mark the center frequencies of the transitions obtained from our simple model. $\nu_0 = 294\,000\,\text{GHz}$.}
    \label{fig:PA}
\end{figure*}

\subsection{Relative line strengths}

We now investigate whether we can use our model to also describe the relative strengths and widths of the resonance lines in the observed data. For this, we consider the measured photoassociation spectra of the $\oneG$ and $\zeroG$ manifolds of \cstate, $\vib'=13$ shown in Fig.$\:$\ref{fig:PA}. These data scans were taken for a constant laser intensity and pulse length and thus the shapes of the lines can be directly compared to each other. Photoassociation loss is governed by the differential equation $\dot{n} = - n^2 \gamma$, where $n$ is the atomic density. $\gamma$ is the rate coefficient which is a function of the laser intensity $I_\mathrm{L}$, the laser detuning $(\nu - \nu_{0,i})$ from each line $i$, and the dipole matrix elements $|M_i|^2 $,
\begin{equation}
    \gamma = \sum_i \gamma_i = \tilde{q}  I_L \sum_i { |M_i|^2 \over 1 + 4 ((\nu-\nu_{0,i})/\Gamma)^2},
    \label{eq:gamma}
\end{equation}
where we use the fact that no saturation effects are  present. In Eq.$\:$(\ref{eq:gamma}), $\tilde{q}$ is an appropriate proportionality constant which we use as a free fit parameter. For the linewidth $\Gamma$ we assume $\Gamma = 2\pi\times12\:\textrm{MHz}$. Our model can be applied to calculate relative values of the dipole matrix elements $|M_i|^2 $ between all excited levels and the initial level depending only on the angular momenta. A Franck-Condon factor which would take into account the overlap of the wave functions of the atomic scattering state and of the excited molecular bound state would be only needed for comparing the transition strengths towards different vibrational levels. In the following, we consider the Franck-Condon factor to be included in the constant $\tilde{q}$. In order to calculate $|M_i|^2 $ we express the scattering state wave function as the symmetrized product wave function of two individual ground state atoms which collide in a  $s$-wave. One of the atoms undergoes an electrical dipole transition, such that the $s$-orbital of the valence electron becomes a $p$-orbital. For driving the transitions we have approximately equal amounts of  $\pi$- and $\sigma$-polarized light. Indeed, when we fit our model to the spectra for an equal mix of polarizations we obtain quite good agreement, see Fig.~\ref{fig:PA}. The only fit parameter for this description is the common constant $\tilde{q}$ which scales the absolute depths and effective widths of the lines. The good agreement between data and fit highlights again the validity of the simple model.

\section{Coupled-channel calculations}
\label{sec:ccModel}
\begin{figure}
	\includegraphics[width=\columnwidth]{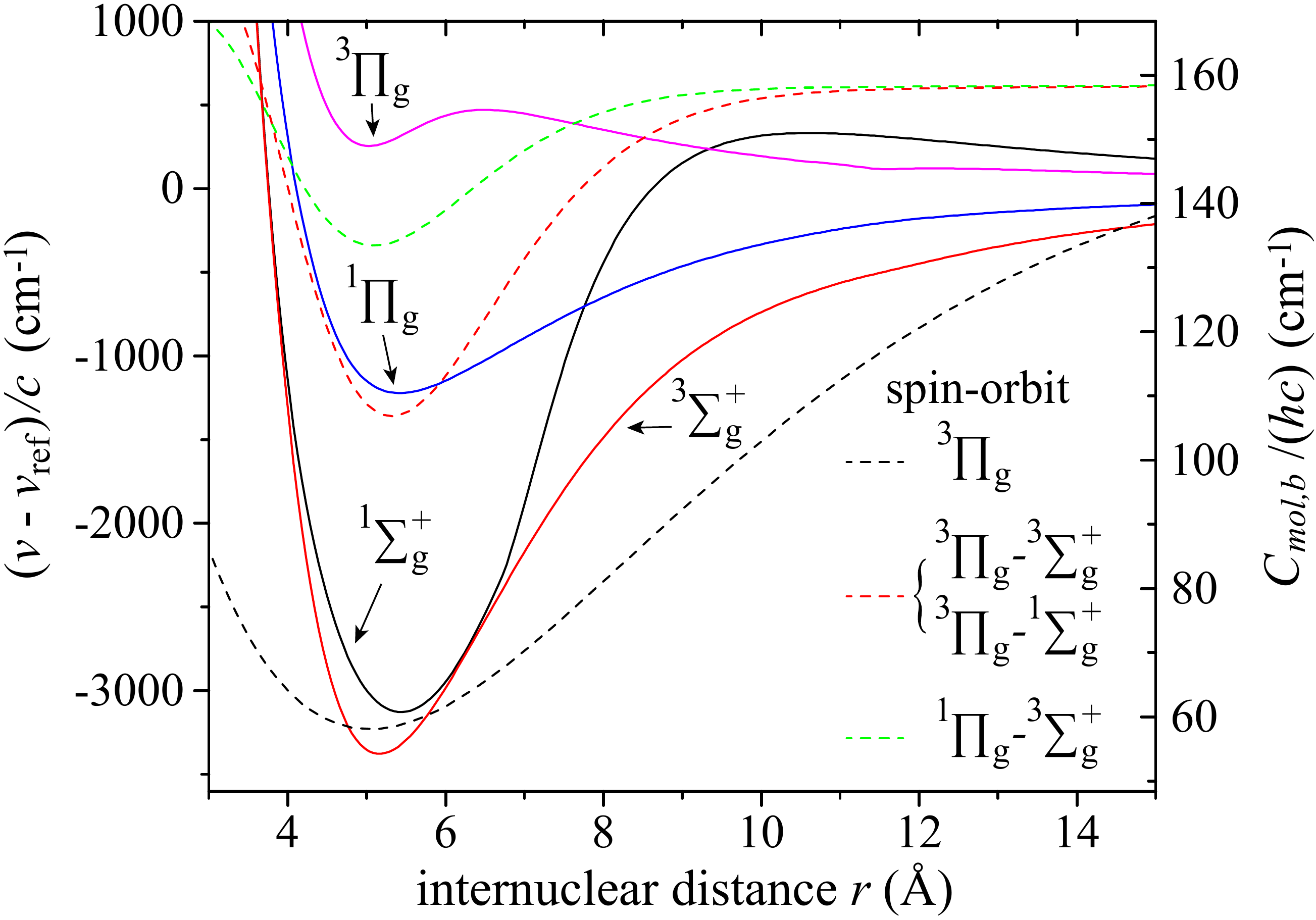}
	\caption{Potential scheme (solid lines) of molecular states with $g$-symmetry at the $s+p$ atom pair asymptote and their constructed spin-orbit coupling $C_{mol,b}$ as function of the internuclear distance $r$ (dashed lines).  $\nu_\text{ref}/c=12737.6\:\textrm{cm}^{-1}$ is the term energy of the $5s + 5p $ atomic pair asymptote.}
	\label{fig:sp}
\end{figure}

So far, we used an isolated electronic state \cstate for our analysis and introduced effective parameters for the different interactions like the spin-spin and hyperfine interaction. Now we extend our view by taking neighboring electronic states into account. This will help us to interpret the derived parameters and the underlying coupling to other states. The relevant molecular states correlate to the atomic pair asymptote $5s \ ^2S + 5p \ ^2P$ and contain sets of electronic $u/g$-symmetry of $^1\Sigma^+, ^3\Sigma^+, ^1\Pi$ and $^3\Pi$. The main coupling between these states is the spin-orbit interaction which gives rise to a splitting of about $237.5\:\textrm{cm}^{-1}$
for the atomic states $^2P_{1/2}$ and $^2P_{3/2}$. Because we want to describe level spectra with an accuracy on the order of $0.001\:\textrm{cm}^{-1}$ this coupling is still relevant even if electronic states are separated by several $1000\:\textrm{cm}^{-1}$. The rotational coupling of electronic states with the selection rule $\Delta \Omega =\pm1$ is automatically included in the Hamiltonian as given below in Eq.$\:$(\ref{eq:HH_CC}). Here, the quantum number $\Omega$ denotes the projection of $\vec{J}$ on the internuclear axis. However, the rotational coupling is of minor importance in our case, as our new data set only contains low rotational levels. Hyperfine coupling breaks the electronic symmetry $u/g$, but in a first step we restrict the model to the $g$-symmetry. In the final calculation we verify that the influence of the neglected states with $u$-symmetry is rather small.

The Hamiltonian for the coupled system for an atom pair $a(^2S)+b(^2P)$ is
\begin{eqnarray}
	\label{eq:ham}
	H  &=& T_n+\sum_i U_i(r)P_i\nonumber\\
	& &+ H_{SO, a} + H_{SO, b}\nonumber\\
	& &+H_{hfs, a} + H_{hfs, b},\
	\label{eq:HH_CC}
\end{eqnarray}
where $T_n$ is the kinetic energy of radial motion and rotation of the pair. $U_i(r)$ denotes the potential energy of the molecular state $i$ and $P_i$  is the  projection operator onto this state. We  sum over all states under consideration. The second and third line give the spin-orbit and hyperfine interactions for atoms $a$ and $b$, respectively.  These  are diagonal in the atom pair basis, conventionally called Hund's case (e), whereas the potential part is diagonal in the Hund's case (a) or (b) basis, specifying the molecular state by the quantum numbers $^{2S+1}\Lambda$.

For setting up the potential curves of the molecular states we use spectroscopic data from the following references. The state $^1\Sigma^+_g$ was studied by Amiot and Verges \cite{Amiot1987,Amiot1990}, $^1\Pi_g$ by Amiot \cite{Amiot1986,Amiot1990} and $^3\Pi_g$ by Bellos \textit{et al.} \cite{Bellos2011}. Data for $^3\Sigma^+_g$ are known from our present work for deeply bound levels and from the work of Tsai \emph{et al.} \cite{Tsai2013} for weakly bound levels. The latter ones were not incorporated in the present analysis, because the energy gap between our data and the data of Ref. \cite{Tsai2013} is too large. A potential construction would then contain much ambiguity, as was also noted in \cite{Tsai2013}. Figure $\:$\ref{fig:sp} shows an overview of the potential scheme. We construct from the spectroscopic work \cite{Amiot1990} a data set for $^1\Sigma^+_g$ and $^1\Pi_g$ which is used in a coupled-channel fit together with our data for \cstate. In Fig.$\:$\ref{fig:sp} one finds also the potential form of the high lying state $^3\Pi_g$, which we derived in a separate analysis to represent approximately  spectroscopic observations of \cite{Bellos2011}. However, we note that a  $^3\Pi_g$ potential curve of high precision is not needed for obtaining reliable results for the low lying levels of \cstate far away from $^3\Pi_g$.

Besides the potential energy curves, Fig.$\:$\ref{fig:sp} also shows the constructed, $r$-dependent spin-orbit interaction which, in general, is different for each molecular state coupling (dashed lines). In the limit $r\rightarrow \infty$, however, it has to converge to the respective atomic value. In the Hamiltonian of Eq.$\:$(\ref{eq:HH_CC}), spin-orbit coupling is dominated  by the part belonging to atom $b(^2P)$, given by $H_{SO,b}=C_{mol,b}\ \vec{S}_b \cdot \vec{L}_b/\hbar^2$.  Here $C_{mol,b}$ is the spin-orbit coupling parameter, which for $r\rightarrow \infty$ reaches the atomic value of $C_{atom,b}=158.3989389\:\textrm{cm}^{-1}\times hc$. The main contribution to the spin-orbit splitting of \cstate comes from the states $^1\Pi_g$ and $^3\Pi_g$ in terms of three spin-orbit functions: a diagonal one for $^3\Pi_g$ and two non-diagonal ones for the couplings of \cstate to $^3\Pi_g$ and to $^1\Pi_g$. In addition, we consider the spin-orbit coupling of $^3\Pi_g$ and $^1\Sigma^+_g$, which, however, influences the levels of \cstate only indirectly and weakly. For the spin-orbit coupling we choose the same functional form as used in \cite{Drozdova2013} for the study of the molecular states of $u$-symmetry in Rb$_2$. This simplifies the later addition of all $u$-states for checking the $u/g$-coupling by the hyperfine interaction. The effective spin-orbit splitting of $^3\Pi_g$ around the potential minimum is only about $58\:\textrm{cm}^{-1}$ (see Fig.$\:$\ref{fig:sp}), as derived from the spectra in \cite{Bellos2011}.

We cannot treat the Zeeman effect in our existing coupled-channel code. The number of channels in the model would become too large because the total angular momentum is no longer a good quantum number. Therefore, we only use the measurements with low magnetic field ($5\:\textrm{G}$) for the analysis.
The experimental data show that the Zeeman shift at low field is smaller than $2\:\textrm{MHz}/\textrm{G}$. In particular,   for the component $0^-_g$ it will be even significantly smaller. Thus, in the worst case we introduce a systematic error of less than $10\:\textrm{MHz}$ which is acceptable for our purposes.

In a first step towards the goal of determining the parameters of the coupled-channel Hamiltonian we approximately remove the hyperfine structure from our data, since the spectroscopic data of the other references do not resolve hyperfine structure. For removing the hyperfine energy we carry out two coupled-channel calculations using the best potentials known at this step. In the first calculation we set the hyperfine interaction to zero and in the second one to the atomic value. The differences of the level positions of these two calculations are the hyperfine splittings. After assigning the observed transitions to the hyperfine levels we subtract their hyperfine shift from the measured transition lines.

We  fit  the potential curves for \cstate, $^1\Pi_g$ and $^1\Sigma^+_g$, and the spin-orbit interaction to the complete data set which exhibits uncertainties between a few to $300\:\textrm{MHz}$. The fit results to a normalized standard deviation of $\sigma= 0.8$ which is quite satisfactory for data with this wide spread of uncertainties. Further information on the representations of the potentials are given in part B of the Supplemental Material, where the respective parameters are listed in Tabs.$\:$\ref{pot1}-\ref{pot4}.

With these results we perform a second fit which now includes hyperfine interaction, e.g.  $H_{hfs, a/b} = a_{mol_{a/b}} \ i_{a/b} \cdot j_{a/b} $, where $i_{a/b}$ and $ j_{a/b}$ are the nuclear spin and the total electronic angular momentum of atom $a$ and atom $b$, respectively. For this fit, we use our primary data with hyperfine structure. The hyperfine parameter $a_{mol_{a/b}}$ for each atom within the molecule is a function of nuclear separation $r$. It is approximated by
\begin{equation}
	a_{mol_{a/b}} = a_{atom_{a/b}} \left(1+\frac{\textit{q}_{a/b}}{\exp((r-r_{hfs})/\Delta r)+1}\right),
	\label{eq:hfs}
\end{equation}
and for $r \rightarrow \infty$ goes over to the respective atomic value $a_{atom_{a/b}}$ \cite{Ye1996, Banerjee2004}. The expression in parentheses is a correction function in order to take into account the changes of the electron distribution due to molecular binding. $\textit{q}_{a/b}$ indicates the fraction by how much the molecular hyperfine splitting deviates from the atomic value. The denominator can be viewed as a switching function around the position $r_{hfs}$ with a width of $\Delta r$. $\textit{q}_{a/b}$ is varied in the fit while the other parameters are arbitrarily set to $r_{hfs}=20\:a_0$ and $\Delta r=2.5\:a_0$ for a reasonably smooth change from the atom pair case at large $r$ to the molecular case around the minimum  of the \cstate potential. Here, $a_0$ denotes the Bohr radius.

Afterwards, several iteration loops are performed for the \cstate potential curve, each followed by a hyperfine fit. For each loop an improved hyperfine correction of the primary data is obtained.  After a few iterations convergence is achieved. All observed levels of \cstate are reproduced with an average deviation of $\pm 9\:\textrm{MHz}$  which is close to the experimental uncertainty of a few MHz and also to the fit quality with the simple model in section \ref{ssec:results}. Hyperfine splittings are, however, often reproduced better. In order to keep the fits meaningful, we restrict the number of fit parameters in these loops as much as possible. Concretely, we use fit parameters to describe the potential minimum of \cstate,  its spin-orbit coupling strength to state $^1\Pi_g$, and the parameter $\textit{q}_{a/b}$ for the hyperfine structure [see Eq.$\:$(\ref{eq:hfs})]. The other spin-orbit functions are used as adjusted during the first step of the set up of our potential scheme, where hyperfine structure was ignored.

From the model at hand, we find that the splitting of about $40\:\textrm{GHz}$ between $0^-_g$ and $1_g$ of \cstate is due to the spin-orbit coupling to $^1\Pi_g$ and $^3\Pi_g$. The admixture of these states to \cstate is around 0.1\% and 0.02\%, respectively.

For atom $a$ (mainly $5s$ character) the molecular hyperfine coupling is about 15.5\% smaller than for the atomic one, i.e. $\textit{q}_a=-0.155$, which probably originates from the lower spin density at the nucleus for the $s$-orbital due to chemical bonding. This suppression is of  similar magnitude as for the triplet ground state of Rb$_2$ \cite{Strauss2010}. For the $p$-electron with its small hyperfine parameter we derive an increase,  $\textit{q}_b=+0.36$. The fit shows a strong anti-correlation between $\textit{q}_a$ and $\textit{q}_b$, resulting in large error limits of about 20\% for $\textit{q}_a$ and of about 50\% for $\textit{q}_b$. The significant increase in $\textit{q}_b$ could be taken as the counterpart of the decrease in $\textit{q}_a$, indicating that the chemical bond  leads to $sp$-hybridization.

Finally, we add to the model all molecular states with $u$-symmetry of the asymptote $5s \ ^2S + 5p \ ^2P$, because the hyperfine interaction  breaks $u/g$-symmetry. For these calculations we use the results on the $u$-states from \cite{Drozdova2013} and on the hyperfine structure of $b^3\Pi_u$ from our work \cite{Deiss2015}. The $u/g$-coupling leads  to an uniform upward shift of the \cstate levels  by not more than $0.1\:\textrm{MHz}$, which can be compensated by lowering the uncoupled potential of state \cstate by the same amount. Additionally, the hyperfine splitting changes by less than 1$\:$MHz, depending on the specific quantum numbers. The admixture of $u$-states to the $g$-states is always less than 0.00001\%. Thus, we conclude that this interaction is unimportant at the present level of our investigation.

\begin{table}[]
		\vspace{-6.5pt}
			\caption{Results of the fits of the simple effective model to the hyperfine level energies obtained from the coupled-channel calculations for the investigated vibrational states $\vib'=0$, 7, 13. All units are in $\text{MHz}{\times}h$. The uncertainties are extracted from the covariance matrix of the fit.}
			\label{tab:resultsccfit}
	\vspace{5pt}
	\centering
		\begin{tabular}{l rl @{~~} rl @{~~} rl}
			\toprule
			\toprule
			& \multicolumn{2}{c}{$\vib'=0$} & \multicolumn{2}{c}{$\vib'=7$} & \multicolumn{2}{c}{$\vib'=13$} \\ \midrule
			$\tilde B_{\vib'}$:   & 431.2           & $\pm$ 0.2       & 419.2           & $\pm$ 0.2       & 409.3           & $\pm$ 0.2        \\
			$\tilde b_{F}$:       & 725.1           & $\pm$ 3.1       & 724.9           & $\pm$ 3.2      & 724.6           & $\pm$ 3.2        \\
			$\tilde c$:           & 110.1           & $\pm$ 5.6      & 110.5           & $\pm$ 5.5      & 110.9           & $\pm$ 5.5       \\
			$\tilde \lambda_{\text{v}'}$:     & 19\,214       & $\pm$ 2       & 20\,356       & $\pm$ 2       & 21\,448       & $\pm$ 2           \\ \bottomrule
			\bottomrule
		\end{tabular}
\end{table}

In order to check for consistency of our two approaches for the data analysis we directly fit our model from section \ref{sec:modelCalc} to the level energies obtained from the coupled-channel calculations. The fits are performed individually for each vibrational manifold, where we restrict the quantum numbers $F' \le 5$ and $J' \le 4$. From the fits the model parameters $\tilde B_{\vib'}$, $\tilde b_F$, $\tilde c$ and $\tilde \lambda$ are extracted, where the tilde symbol distinguishes them from the parameters corresponding to the experimental data. The results are given in table \ref{tab:resultsccfit}. The rms deviations of these fits are about 7$\:$MHz for all levels. These deviations as well as the fit uncertainties in table \ref{tab:resultsccfit} indicate fortunately a comparatively small difference between the two approaches, on the order of the experimental uncertainty. This allows for extracting interesting information from the coupled-channel model, e.g. that we can expect the hyperfine coupling constants $\tilde b_{F}$ $(b_{F})$ and  $\tilde c$ $(c)$ to be rather constant over a large vibrational range. As another check that the effective and the coupled-channel model are compatible with each other at the given level of precision, we compare table \ref{tab:results} with table \ref{tab:resultsccfit}. Indeed, all molecule parameter values for the rotational constant $\tilde B_{\vib'}$ $(B_{\vib'})$, the effective spin-spin coupling $\tilde \lambda$ $(\lambda)$, and for the hyperfine constants $\tilde b_{F}$ $(b_{F})$ and  $\tilde c$ $(c)$ agree well within the fit uncertainties.

\section{Conclusion}
\label{sec:conclusion}
We experimentally investigate the hyperfine structures of the vibrational levels $\vib'=0$, 7, 13 of the \cstate potential. The data is obtained from photoassociation and photoexcitation spectroscopy at different magnetic fields starting from various precisely-defined initial states. Using a simple model, we unambiguously reproduce the measured transition lines both in frequency position and line strength. The experimental identification of hyperfine levels with nuclear spin quantum number $I'=1$ by selective state preparation reveals a wrong assignment of some spectroscopic resonances in our previous work~\cite{Takekoshi2011}. From fits of our model to the data we extract the relevant molecular parameters individually for each observed vibrational level with low uncertainties. In contrast to our former investigation \cite{Takekoshi2011} the anisotropic hyperfine constants are also determined.

The fit of potentials and spin-orbit couplings with the help of coupled-channel calculations is a simultaneous representation of all observations and allows for identifying clearly the origin of the spin-spin interaction introduced for the simple model evaluation as spin-orbit interaction with $^1\Pi_g$ and $^3\Pi_g$. Furthermore, we find that the observed hyperfine splitting can be quite well described by using an atomic hyperfine interaction of the $s$-electron when reducing it by 15.5\%. A decrease by about 7\% of the hyperfine parameter is reported in \cite{Tsai2013} for data on weakly bound levels of state 1$_g$ correlated to \cstate. The difference between deeply and weakly bound levels could be related to the variation of spin density by the chemical bond. The increase of the hyperfine interaction by the $p$-electron shows the counterpart resulting from the distorted $s$-electron.

In general the simple model can reproduce quite well the experimental data as well as the coupled-channel calculations with an accuracy down to several MHz. However, on that level of accuracy systematic deviations between the models start to appear. An extension of the analysis is possible once more measured data on the state \cstate are available. The coupled-channel calculations presented here can be used to predict the level energies with a precision of a few tens of MHz for the range between $\vib'= 0$  to $\vib'=13$ and low rotational states $J'\leq 4$.

\begin{acknowledgments}
This work is funded by the German Research Foundation (DFG). B.D., M.D., J.W., and J.H.D. would like to thank the BaRbI team for support.  E.T. gratefully acknowledges support from the Minister of Science and Culture of Lower Saxony, Germany, by providing a Niedersachsenprofessur.
\end{acknowledgments}

\bibliographystyle{apsprl}
\newpage

\section{Supplemental Material}
\subsection{Spectroscopic data}
\begin{table}[h!]
			\vspace{-6.5pt}
	   \caption{Dark-state spectroscopy results of \astate, $\vib=0$ levels as shown in Fig.~\ref{fig:groundstate}, with corresponding quantum numbers. The measurements are taken at a $B$-field of 999.9 G.}
	   \label{darkstate}
	 \vspace{5pt}
    \centering
    \begin{tabular}{c @{\hspace{5mm}}C{3mm}C{3mm}C{3mm}C{3mm}}
    	\toprule
    	\toprule
    	\centering $\nu_\mathrm{meas}$ & $f$ & $F$ & $I$ & $R$ \\
    	       \centering (GHz)        &     &     &     &     \\ \midrule
    	          -7024.144            & 4   &  4  &  3  &  0  \\
    	          -7024.436            & 2   &  3  &  1  &  2  \\
    	          -7034.138            & 0   &  2  &  1  &  2  \\
    	          -7037.857            & 2   &  2  &  3  &  2  \\
    	          -7039.787            & 2   &  2  &  3  &  0  \\ \bottomrule
    	          \bottomrule
    \end{tabular}
\end{table}
\newpage

\begin{table}[h!]
			\vspace{-6.5pt}
	   \caption{Measured term frequencies $\nu_\mathrm{meas}-v_0$ for the $\vib' = 0$ manifold with ${\nu_0=281\,000\,\text{GHz}}$. $\Delta \nu$ gives the difference between the data and the fitted model calculations, based on the parameters listed in Tab.~\ref{tab:results}. In addition, the initial state of the spectroscopy, the polarization (pol.) of the light and the resulting quantum numbers are given for each measurement. The root mean square error (rmse) of the fit is 11.4\,MHz.
}\label{tab:v0}
	   \vspace{5pt}
    \centering
    \begin{tabular}{L{4mm}@{\hspace{-2mm}}R{14mm}C{8mm}@{\hspace{4mm}}L{11mm}@{\hspace{-1mm}}C{10mm}@{\hspace{3mm}}R{3mm}R{5mm}R{5mm}R{4mm}}
        \toprule
        \toprule
        & \centering $\nu_\mathrm{meas}-v_0$ & $\Delta\nu$ & initial & pol.& $I'$ & $J'$ & $F'$ & $m'_\mathrm{F}$ \\
        & \centering \footnotesize (GHz)               & \footnotesize (MHz) & state   &     &      &      &      &         \\ \midrule
        \multicolumn{9}{c}{$B = 5$\,G}                                                     \\ \midrule
        \oneG
        & 71.282  & -12 & $|a$,\,$R$=0$\rangle$ & $\pi$        & 3 & 1.2 & 2.0 & +2 \\
        & 71.287  & -1  & PA                    & $\pi+\sigma$ & 3 & 1.2 & 2.0 & -2 \\
        & 71.986  & +6  & PA                    & $\pi+\sigma$ & 3 & 1.3 & 3.0 & -1 \\
        & 73.370  & +0  & PA                    & $\pi+\sigma$ & 3 & 2.0 & 1.0 & -1 \\
        & 73.936  & -4  & PA                    & $\pi+\sigma$ & 3 & 2.0 & 2.0 & -1 \\
        & 74.846  & +1  & PA                    & $\pi+\sigma$ & 1 & 1.9 & 2.0 & -2 \\
        & 75.033  & +0  & PA                    & $\pi+\sigma$ & 1 & 2.1 & 3.0 & -3 \\ \midrule
        \zeroG & 111.787 & -18 & $|a$,\,$R$=0$\rangle$ & $\pi$        & 3 & 0.0 & 3.0 & +2 \\
        & 114.329 & -22 & $|a$,\,$F$=4$\rangle$ & $\pi$        & 3 & 2.0 & 5.0 & +4 \\
        & 114.426 & -11 & $|a$,\,$F$=4$\rangle$ & $\pi$        & 3 & 2.0 & 4.0 & +4 \\
        & 114.548 & -14 & $|a$,\,$F$=4$\rangle$ & $\sigma$     & 3 & 2.0 & 3.0 & +3 \\
        & 114.549 & -14 & $|a$,\,$R$=0$\rangle$ & $\pi$        & 3 & 2.0 & 3.0 & +2 \\
        & 114.660 & -17 & $|a$,\,$R$=0$\rangle$ & $\pi$        & 3 & 2.0 & 2.0 & +2 \\
        & 114.749 & -24 & $|a$,\,$R$=0$\rangle$ & $\sigma$     & 3 & 2.0 & 1.0 & +1 \\
        & 120.757 & +3  & $|a$,\,$F$=4$\rangle$ & $\pi$        & 3 & 4.0 & 5.0 & +4 \\
        & 120.884 & +23 & $|a$,\,$F$=4$\rangle$ & $\pi$        & 3 & 4.0 & 4.0 & +4 \\ \midrule
        \multicolumn{9}{c}{$B = 250$\,G}                                                     \\ \midrule
        \oneG  & 70.988  & -1  & $|a$,\,$R$=0$\rangle$ & $\sigma$     & 3 & 1.2 & 2.1 & +1 \\
        \zeroG & 111.817 & -9  & $|a$,\,$R$=0$\rangle$ & $\pi$        & 3 & 0.0 & 3.0 & +2 \\ \midrule
        \multicolumn{9}{c}{$B = 500$\,G}                                                     \\ \midrule
        \oneG  & 70.580  & -4  & $|a$,\,$R$=0$\rangle$ & $\sigma$     & 3 & 1.3 & 2.2 & +1 \\
        \zeroG & 111.768 & -11 & $|a$,\,$I$=1$\rangle$ & $\sigma$     & 1 & 0.0 & 1.0 & +1 \\
        & 111.866 & -2  & $|a$,\,$R$=0$\rangle$ & $\pi$        & 3 & 0.0 & 3.0 & +2 \\ \midrule
        \multicolumn{9}{c}{$B = 750$\,G}                                                     \\ \midrule
        \oneG  & 70.109  & -9  & $|a$,\,$R$=0$\rangle$ & $\sigma$     & 3 & 1.3 & 2.3 & +1 \\
        \zeroG & 111.944 & -7  & $|a$,\,$R$=0$\rangle$ & $\pi$        & 3 & 0.0 & 3.0 & +2 \\ \midrule
        \multicolumn{9}{c}{$B = 999.9$\,G}                                                     \\ \midrule
        \oneG  & 69.213  & -17 & FB                    & $\pi$        & 3 & 1.4 & 2.2 & +2 \\
        & 69.593  & -8  & $|a$,\,$R$=0$\rangle$ & $\sigma$     & 3 & 1.3 & 2.4 & +1 \\
        & 70.915  & -13 & FB                    & $\pi$        & 3 & 1.6 & 3.2 & +2 \\
        & 73.279  & -10 & FB                    & $\pi$        & 3 & 2.3 & 2.5 & +2 \\
        & 73.420  & -4  & FB                    & $\pi$        & 1 & 1.8 & 2.3 & +2 \\ \midrule
        \zeroG & 111.866 & +22 & $|a$,\,$I$=1$\rangle$ & $\sigma$     & 1 & 0.0 & 1.0 & +1 \\
        & 112.024 & +1  & FB                    & $\pi$        & 3 & 0.0 & 3.0 & +2 \\
        & 112.027 & -2  & $|a$,\,$R$=0$\rangle$ & $\pi$        & 3 & 0.0 & 3.0 & +2 \\
        & 114.310 & -1  & FB                    & $\pi$        & 1 & 2.0 & 2.6 & +2 \\
        & 114.410 & +6  & $|a$,\,$I$=1$\rangle$ & $\pi$        & 1 & 2.0 & 2.5 & +2 \\
        & 114.416 & +0  & FB                    & $\pi$        & 1 & 2.0 & 2.5 & +2 \\
        & 114.468 & +3  & FB                    & $\pi$        & 3 & 2.0 & 4.1 & +2 \\
        & 114.545 & +4  & $|a$,\,$I$=1$\rangle$ & $\sigma$     & 1 & 2.0 & 2.0 & +1 \\
        & 114.668 & +7  & FB                    & $\pi$        & 3 & 2.0 & 3.4 & +2 \\
        & 114.964 & +20 & FB                    & $\pi$        & 3 & 2.0 & 2.5 & +2 \\ \bottomrule
        \bottomrule
    \end{tabular}
\end{table}
\newpage

\begin{table}[h!]
			\vspace{-6.5pt}
		\caption{Measured term frequencies $\nu_\mathrm{meas}-v_0$ for the $\vib' = 7$ manifold with ${\nu_0=288\,000\,\text{GHz}}$. $\Delta \nu$ gives the difference between the data and the fitted model calculations, based on the parameters listed in Tab.~\ref{tab:results}. In addition, the initial state of the spectroscopy, the polarization (pol.) of the light and the resulting quantum numbers are given for each measurement. The root mean square error (rmse) of the fit is 4.3\,MHz.}\label{tab:v7}
		\vspace{5pt}
	\centering
	 \begin{tabular}{L{4mm}@{\hspace{-2mm}}R{14mm}C{8mm}@{\hspace{4mm}}L{11mm}@{\hspace{-1mm}}C{10mm}@{\hspace{3mm}}R{3mm}R{5mm}R{5mm}R{4mm}}
		\toprule
		\toprule
		& \centering$\nu_\mathrm{meas}-\nu_0$ & $\Delta\nu$ & initial & pol.& $I'$ & $J'$ & $F'$ & $m'_\mathrm{F}$ \\
		& \centering \footnotesize(GHz)               & \footnotesize(MHz)       & state   &     &      &      &      &     \\ \midrule
		\multicolumn{9}{c}{$B = 5$\,G}                                             \\ \midrule
		\oneG  & 497.551 & +3 & PA & $\pi+\sigma$ & 3 & 1.2 & 2.0 & -1 \\
		& 498.256 & +3 & PA & $\pi+\sigma$ & 3 & 1.4 & 3.0 & -1 \\
		& 499.627 & -6 & PA & $\pi+\sigma$ & 3 & 2.0 & 1.0 & -1 \\
		& 500.179 & -3 & PA & $\pi+\sigma$ & 3 & 2.0 & 2.0 & -1 \\
		& 501.086 & +4 & PA & $\pi+\sigma$ & 1 & 1.9 & 2.0 & -2 \\
		& 503.021 & -1 & PA & $\pi+\sigma$ & 3 & 3.0 & 2.0 & -1 \\ \midrule
		\zeroG & 540.325 & -4 & PA & $\pi+\sigma$ & 3 & 0.0 & 3.0 & -1 \\
		& 542.810 & +4 & PA & $\pi+\sigma$ & 1 & 2.0 & 3.0 & -1 \\
		& 543.003 & +7 & PA & $\pi+\sigma$ & 3 & 2.0 & 3.0 & -3 \\
		& 543.113 & +0 & PA & $\pi+\sigma$ & 3 & 2.0 & 2.0 & -2 \\
		& 543.198 & -7 & PA & $\pi+\sigma$ & 3 & 2.0 & 1.0 & -1 \\ \bottomrule
		\bottomrule
	\end{tabular}
\end{table}

\begingroup
	\squeezetable
\begin{table}
			\vspace{-6.5pt}
	    \caption{Measured term frequencies $\nu_\mathrm{meas}-v_0$ for the $\vib' = 13$ manifold with ${\nu_0=294\,000\,\text{GHz}}$. $\Delta \nu$ gives the difference between the data and the fitted model calculations, based on the parameters listed in Tab.~\ref{tab:results}. In addition, the initial state of the spectroscopy, the polarization (pol.) of the light and the resulting quantum numbers are given for each measurement. The root mean square error (rmse) of the fit is 13.9\,MHz.}\label{tab:v13a}
	    \vspace{5pt}
    \centering
    \begin{tabular}{L{4mm}@{\hspace{-2mm}}R{14mm}C{8mm}@{\hspace{4mm}}L{11mm}@{\hspace{-1mm}}C{10mm}@{\hspace{3mm}}R{3mm}R{5mm}R{5mm}R{4mm}}
        \toprule
        \toprule
        &\centering $\nu_\mathrm{meas}-\nu_0$ & $\Delta\nu$ & initial & pol.& $I'$ & $J'$ & $F'$ & $m'_\mathrm{F}$ \\
        &\centering (GHz)               & (MHz)       & state   &     &      &      &      &  \\ \midrule
        \multicolumn{9}{c}{$B = 5$\,G}                                             \\ \midrule
        \oneG  & 626.527 & -8  & $|a$,\,$R$=0$\rangle$ &    $\pi$     & 3 & 1.2 & 2.0 & +2 \\
        & 626.532 & -0  &          PA           & $\pi+\sigma$ & 3 & 1.2 & 2.0 & -1 \\
        & 627.234 & -1  & $|a$,\,$R$=0$\rangle$ &    $\pi$     & 3 & 1.4 & 3.0 & +2 \\
        & 627.235 & +2  &          PA           & $\pi+\sigma$ & 3 & 1.4 & 3.0 & -1 \\
        & 627.960 & -2  &          PA           & $\pi+\sigma$ & 1 & 1.1 & 1.0 & -1 \\
        & 628.592 & -9  & $|a$,\,$R$=0$\rangle$ &   $\sigma$   & 3 & 2.0 & 1.0 & +1 \\
        & 628.593 & -1  &          PA           & $\pi+\sigma$ & 3 & 2.0 & 1.0 & -1 \\
        & 629.136 & -6  &          PA           & $\pi+\sigma$ & 3 & 2.0 & 2.0 & -1 \\
        & 629.137 & -6  & $|a$,\,$R$=0$\rangle$ &    $\pi$     & 3 & 2.0 & 2.0 & +2 \\
        & 629.651 & -4  &          PA           & $\pi+\sigma$ & 1 & 1.9 & 1.0 & -1 \\
        & 629.779 & +0  &          PA           & $\pi+\sigma$ & 3 & 2.0 & 3.0 & -3 \\
        & 630.045 & -3  &          PA           & $\pi+\sigma$ & 1 & 1.9 & 2.0 & -1 \\
        & 630.206 & -0  &          PA           & $\pi+\sigma$ & 1 & 2.1 & 3.0 & -1 \\
        & 631.821 & -6  &          PA           & $\pi+\sigma$ & 3 & 3.0 & 1.0 & -1 \\
        & 631.915 & -0  &          PA           & $\pi+\sigma$ & 3 & 3.0 & 2.0 & -2 \\
        & 632.146 & +4  &          PA           & $\pi+\sigma$ & 3 & 3.0 & 3.0 & -3 \\
        & 632.540 & -4  &          PA           & $\pi+\sigma$ & 1 & 3.0 & 3.0 & -1 \\ \midrule
        \zeroG & 671.411 & -7  &          PA           & $\pi+\sigma$ & 1 & 0.0 & 1.0 & -1 \\
        & 671.489 & -15 &          PA           & $\pi+\sigma$ & 3 & 0.0 & 3.0 & -1 \\
        & 671.490 & -16 & $|a$,\,$R$=0$\rangle$ &    $\pi$     & 3 & 0.0 & 3.0 & +2 \\
        & 673.905 & -10 & $|a$,\,$F$=4$\rangle$ &    $\pi$     & 3 & 2.0 & 5.0 & +4 \\
        & 673.911 & +0  &          PA           & $\pi+\sigma$ & 1 & 2.0 & 3.0 & -1 \\
        & 673.986 & -1  &          PA           & $\pi+\sigma$ & 3 & 2.0 & 4.0 & -3 \\
        & 673.994 & -9  & $|a$,\,$F$=4$\rangle$ &    $\pi$     & 3 & 2.0 & 4.0 & +4 \\
        & 674.092 & -3  &          PA           & $\pi+\sigma$ & 3 & 2.0 & 3.0 & -1 \\
        & 674.096 & -7  & $|a$,\,$R$=0$\rangle$ &    $\pi$     & 3 & 2.0 & 3.0 & +2 \\
        & 674.097 & -8  & $|a$,\,$F$=4$\rangle$ &   $\sigma$   & 3 & 2.0 & 3.0 & +3 \\
        & 674.195 & -11 &          PA           & $\pi+\sigma$ & 3 & 2.0 & 2.0 & -1 \\
        & 674.198 & -13 & $|a$,\,$R$=0$\rangle$ &    $\pi$     & 3 & 2.0 & 2.0 & +2 \\
        & 674.275 & -21 &          PA           & $\pi+\sigma$ & 3 & 2.0 & 1.0 & -1 \\
        & 674.279 & -23 & $|a$,\,$R$=0$\rangle$ &   $\sigma$   & 3 & 2.0 & 1.0 & +1 \\
        & 679.984 & +11 & $|a$,\,$F$=4$\rangle$ &    $\pi$     & 3 & 4.0 & 5.0 & +4 \\
        & 680.117 & +6  & $|a$,\,$F$=4$\rangle$ &    $\pi$     & 3 & 4.0 & 4.0 & +4 \\
        & 680.228 & +6  & $|a$,\,$F$=4$\rangle$ &   $\sigma$   & 3 & 4.0 & 3.0 & +3 \\ \midrule
        \multicolumn{9}{c}{$B = 999.9$\,G}                                             \\ \midrule
        \oneG  & 624.443 & +5  &          FB           &  $\pi$   & 3 & 1.4 & 2.2 & +2 \\
        & 626.138 & +7  &          FB           &  $\pi$   & 3 & 1.6 & 3.2 & +2 \\
        & 626.139 & +6  &          FB           &  $\pi$   & 3 & 1.6 & 3.2 & +2 \\
        & 628.415 & +0  &          FB           &  $\pi$   & 3 & 2.4 & 2.5 & +2 \\
        & 628.416 & -1  &          FB           &  $\pi$   & 3 & 2.4 & 2.5 & +2 \\
        & 628.591 & +15 & $|a$,\,$I$=1$\rangle$ &  $\pi$   & 1 & 1.8 & 2.3 & +2 \\
        & 628.601 & +6  &          FB           &  $\pi$   & 1 & 1.8 & 2.3 & +2 \\
        & 628.602 & +5  &          FB           &  $\pi$   & 1 & 1.8 & 2.3 & +2 \\
        & 629.536 & +26 &          FB           &  $\pi$   & 3 & 2.4 & 3.3 & +2 \\
        & 630.151 & +15 &          FB           &  $\pi$   & 3 & 2.5 & 4.9 & +2 \\
        & 630.466 & +34 &          FB           &  $\pi$   & 1 & 2.3 & 2.9 & +2 \\
        & 630.484 & +15 & $|a$,\,$I$=1$\rangle$ &  $\pi$   & 1 & 2.3 & 2.9 & +2 \\
        & 630.911 & +34 &          FB           &  $\pi$   & 3 & 2.3 & 3.9 & +2 \\
        & 631.500 & +7  &          FB           &  $\pi$   & 3 & 3.0 & 2.7 & +2 \\
        & 632.501 & +21 &          FB           &  $\pi$   & 3 & 2.9 & 3.3 & +2 \\
        & 633.006 & +21 &          FB           &  $\pi$   & 1 & 2.8 & 3.0 & +2 \\
        & 633.026 & +1  & $|a$,\,$I$=1$\rangle$ &  $\pi$   & 1 & 2.8 & 3.0 & +2 \\
        & 633.251 & +15 &          FB           &  $\pi$   & 3 & 2.7 & 3.9 & +2 \\
        & 635.494 & -1  &          FB           &  $\pi$   & 3 & 4.0 & 3.1 & +2 \\
        & 636.106 & -4  &          FB           &  $\pi$   & 1 & 3.9 & 3.2 & +2 \\
        & 636.139 & +5  &          FB           &  $\pi$   & 1 & 4.0 & 4.1 & +2 \\
        & 636.154 & -1  &          FB           &  $\pi$   & 3 & 3.9 & 4.3 & +2 \\ \midrule
        \zeroG & 671.550 & +31 &          FB           & $\sigma$ & 1 & 0.0 & 1.0 & +1 \\
        & 671.682 & +23 &          FB           &  $\pi$   & 3 & 0.0 & 3.0 & +2 \\
        & 671.705 & -0  & $|a$,\,$R$=0$\rangle$ &  $\pi$   & 3 & 0.0 & 3.0 & +2 \\
        & 671.736 & +24 &          FB           & $\sigma$ & 3 & 0.0 & 3.0 & +3 \\
        & 673.882 & +17 &          FB           &  $\pi$   & 1 & 2.0 & 2.6 & +2 \\
        & 673.915 & +3  &          FB           & $\sigma$ & 3 & 2.0 & 4.2 & +1 \\
        & 673.971 & +19 &          FB           &  $\pi$   & 1 & 2.0 & 2.5 & +2 \\
        & 674.023 & +16 &          FB           &  $\pi$   & 3 & 2.0 & 4.1 & +2 \\
        & 674.076 & -4  &          FB           & $\sigma$ & 3 & 2.0 & 4.2 & +3 \\
        & 674.193 & +22 &          FB           &  $\pi$   & 3 & 2.0 & 3.4 & +2 \\
        & 674.241 & +21 &          FB           & $\sigma$ & 3 & 2.0 & 3.3 & +3 \\
        & 674.392 & +24 &          FB           & $\sigma$ & 3 & 2.0 & 2.7 & +1 \\
        & 674.462 & +25 &          FB           &  $\pi$   & 3 & 2.0 & 2.5 & +2 \\
        & 674.757 & +28 &          FB           & $\sigma$ & 3 & 2.0 & 1.5 & +1 \\ \bottomrule
        \bottomrule
    \end{tabular}
\end{table}
\endgroup
\newpage

\subsection{Potential curves}
The potentials $U(r)$  for the states \cstate, $^1\Pi_g$ and $^1\Sigma^+_g$ are represented in an analytic form within three ranges of $r$: the short-range $U_{\rm SR}(r)$, the intermediate-range $U_{\mathrm {IR}}(r)$, and the long-range $U_{\mathrm {LR}}(r)$. For the
intermediate range we use a power expansion \cite{Strauss2010}
\begin{equation}
\label{eq:uanal}
U_{\mathrm {IR}}(r)=\sum_{i=0}^{n}a_i\,\xi^i(r),
\end{equation}
with the expansion variable $\xi$ being non-linear in the nuclear separation $r$,
\begin{equation}
\label{eq:xv}
\xi(r)=\frac{r - r_m}{r + b\,r_m}.
\end{equation}
The coefficients $a_i$ are the fitting parameters.  $r_m$ is typically set close to the minimum of the potential and $b$ is adjusted to best describe the asymmetry of the potential function about the minimum. The potential is extrapolated for $r < r_{\rm SR}$  by the short-range part $U_{\rm SR}$ with
\begin{equation}
\label{eq:rep}
  U_{\rm SR}(r)= u_1 + u_2/r^{N_s},
\end{equation}
where the parameters $u_1$ and $u_2$ are determined such that a continuous and differentiable
connection at $r_{\rm SR}$ with $N_s=6$ is assured. The long-range part ($r > r_{\rm LR}$),
\begin{equation}
\label{eq:lr}
U_{\mathrm {LR}}(r)=-C_3/r^3-C_6/r^6-C_{8}/r^{8}- ...,
\end{equation}
assures the proper long-range behavior towards the atom pair asymptote $5s \ ^2S + 5p \ ^2P$. Because we describe deeply bound levels, the specific form of the long-range part is not relevant for the present analysis, but it would become important if the data from Tsai \emph{et al.} \cite{Tsai2013} were also incorporated. With respect to the ground state asymptote $5s \ ^2S + 5s \ ^2S$, $(f_a=1)+(f_b=1)$ which was chosen as the energy reference in this work, the excited pair asymptote is located at 12579.235953~\wn.

Tables \ref{pot1}, \ref{pot2}, \ref{pot3} give the results for the states \cstate, $^1\Pi_g$, $^1\Sigma^+_g$, respectively.

The energetically high lying excited state $^3\Pi_g$ has a potential maximum at relatively short internuclear distance (see Fig.~\ref{fig:sp}) and thus the analytic form according to Eq.$\:$(\ref{eq:uanal}) is not well suited for its representation. Instead we apply a cubic spline form with a long-range extension as in Eq.$\:$(\ref{eq:lr}). Table \ref{pot4} shows the list of pairs of internuclear separations and potential energies, with which the spline potential is constructed. In addition, the long-range parameters ${C_3}, {C_6}, {C_8} $ are given.

\begin{table}
			\vspace{-6.5pt}
\fontsize{8pt}{13pt}\selectfont \caption{Parameters of the
analytic representation of the potential curve of state \cstate. The
energy reference is the $5s+5p$ dissociation asymptote. Parameters with
$\ast$ are set for continuous extrapolation of the potential and those with $\ast \ast$ are taken from \cite{Tsai2013,Marinescu1995}. }
\label{pot1}
\vspace{5pt}
\begin{tabular*}{0.6\columnwidth}{@{\extracolsep{\fill}}lr}
\hline
\hline
   \multicolumn{2}{c}{$r < r_\mathrm{SR}=$ 3.825 \AA}    \\
\hline
   $u_1^\ast$ & -3.25943420$\times 10^{3}$ \wn \\
   $u_2^\ast$ & 8.35611567$\times 10^{6}$  \wn \AA $^{6}$ \\
\hline
   \multicolumn{2}{c}{$r_\mathrm{SR} \leq r \leq r_\mathrm{LR}=$ 7.000 \AA}    \\
\hline
    $b$ &   $0.0$              \\
    $r_\mathrm{m}$ & 5.164770 \AA               \\
    $a_{0}$ &  -3376.946394 \wn\\
    $a_{1}$ & -1.86075526142358939$\times 10^{ 2}$ \wn\\
    $a_{2}$ &  2.31591930225489923$\times 10^{ 4}$ \wn\\
    $a_{3}$ & -1.56788016077407647$\times 10^{ 4}$ \wn\\
    $a_{4}$ & -2.44553017859753927$\times 10^{ 4}$ \wn\\
    $a_{5}$ &  7.58858636438533431$\times 10^{ 4}$ \wn\\
    $a_{6}$ & -3.79933890563158348$\times 10^{ 4}$ \wn\\
    $a_{7}$ & -5.04140528538642393$\times 10^{ 4}$ \wn\\
    $a_{8}$ &  1.77923885437870456$\times 10^{ 5}$ \wn\\
    $a_{9}$ &  2.29288037865693623$\times 10^{ 5}$ \wn\\
   $a_{10}$ & -3.93764864037126594$\times 10^{ 5}$ \wn\\
\hline
   \multicolumn{2}{c}{$ r > r_\mathrm{LR}$}\\
\hline
 ${C_3}^{\ast\ast}$ &    5.735328470$\times 10^{5}$ \wn\AA$^3$      \\
 ${C_{6}}^{\ast\ast}$ &  2.0221400607$\times 10^{8}$ \wn\AA$^6$   \\
 ${C_{8}}^{\ast\ast}$ & 1.04901144487$\times 10^{9}$ \wn\AA$^{8}$   \\
 ${C_{10}}^{\ast}$ & -9.2704471$\times 10^{11}$ \wn\AA$^{10}$   \\
 ${C_{12}}^{\ast}$ &  2.5729090$\times 10^{13}$ \wn\AA$^{12}$   \\
\hline
\hline
\end{tabular*}
\end{table}

\begin{table}
			\vspace{-6.5pt}
\fontsize{8pt}{13pt}\selectfont \caption{Parameters of the
analytic representation of the potential curve of state $^1\Pi_g$. The
energy reference is the $5s+5p$ dissociation asymptote. Parameters with
$\ast$ are set for continuous extrapolation of the potential and those with $\ast \ast$ are taken from \cite{Tsai2013,Marinescu1995}.}
\label{pot2}
\vspace{5pt}
\begin{tabular*}{0.6\columnwidth}{@{\extracolsep{\fill}}lr}
\hline
\hline
   \multicolumn{2}{c}{$r < r_\mathrm{SR}=$ 4.206 \AA}    \\
\hline
   ${u_1}^\ast$ & -1.6379553$\times 10^{3}$ \wn \\
   ${u_2}^\ast$ &  7.8329961$\times 10^{6}$  \wn \AA $^{6}$ \\
\hline
   \multicolumn{2}{c}{$r_\mathrm{SR} \leq r \leq r_\mathrm{LR}=$ 11.400 \AA}    \\
\hline
    $b$ &   $-0.4$              \\
    $r_\mathrm{m}$ & 5.42860 \AA               \\
    $a_{0}$ &  -1222.30537 \wn\\
    $a_{1}$ &  1.87328847142396100$\times 10^{ 1}$ \wn\\
    $a_{2}$ &  3.27627846134345646$\times 10^{ 3}$ \wn\\
    $a_{3}$ & -2.41144186476804350$\times 10^{ 2}$ \wn\\
    $a_{4}$ & -1.89905524402217793$\times 10^{ 3}$ \wn\\
    $a_{5}$ &  2.01930422858255361$\times 10^{ 3}$ \wn\\
    $a_{6}$ &  1.39513588881434771$\times 10^{ 3}$ \wn\\
    $a_{7}$ & -1.21928763590355520$\times 10^{ 4}$ \wn\\
    $a_{8}$ &  7.16767032715221376$\times 10^{ 3}$ \wn\\
    $a_{9}$ &  3.48980993982300715$\times 10^{ 4}$ \wn\\
   $a_{10}$ & -6.54764720906592702$\times 10^{ 4}$ \wn\\
   $a_{11}$ & -7.48880949630579125$\times 10^{ 4}$ \wn\\
   $a_{12}$ &  1.62482105910330632$\times 10^{ 5}$ \wn\\
   $a_{13}$ &  8.14091568750822335$\times 10^{ 4}$ \wn\\
   $a_{14}$ & -1.51306348841136700$\times 10^{ 5}$ \wn\\
\hline
   \multicolumn{2}{c}{$ r > r_\mathrm{LR}$}\\
\hline
 ${C_3}^{\ast\ast}$ &    2.896150$\times 10^{5}$ \wn\AA$^3$      \\
 ${C_{6}}^{\ast\ast}$ &  3.878000$\times 10^{7}$ \wn\AA$^6$   \\
 ${C_{8}}^{\ast\ast}$ & 5.6722005$\times 10^{8}$ \wn\AA$^{8}$   \\
 ${C_{10}}^{\ast}$ & 4.07340914$\times 10^{11}$ \wn\AA$^{10}$   \\
 ${C_{12}}^{\ast}$ &-3.80261888$\times 10^{13}$ \wn\AA$^{12}$   \\
\hline
\hline
\end{tabular*}
\end{table}

\begin{table}
			\vspace{-6.5pt}
\fontsize{8pt}{13pt}\selectfont \caption{Parameters of the
analytic representation of the potential curve of state $^1\Sigma^+_g$. The
energy reference is the $5s+5p$ dissociation asymptote. Parameters with
$\ast$ are set for continuous extrapolation of the potential and those with $\ast \ast$ are taken from \cite{Tsai2013,Marinescu1995}.}
\label{pot3}
\vspace{5pt}
\begin{tabular*}{0.6\columnwidth}{@{\extracolsep{\fill}}lr}
\hline
\hline
   \multicolumn{2}{c}{$r < r_\mathrm{SR}=$ 4.437 \AA}    \\
\hline
   ${u_1}^\ast$ & -3.703000$\times 10^{3}$ \wn \\
   ${u_2}^\ast$ & 1.05251166$\times 10^{7}$  \wn \AA $^{6}$ \\
\hline
   \multicolumn{2}{c}{$r_\mathrm{SR} \leq r \leq r_\mathrm{LR}=$ 6.74 \AA}    \\
\hline
    $b$ &   $0.0$              \\
    $r_\mathrm{m}$ & 5.43040 \AA               \\
    $a_{0}$ &  -3127.71366 \wn\\
    $a_{1}$ &  -4.39321102104440584$\times 10^{ 1}$ \wn\\
    $a_{2}$ &   1.84395440542794568$\times 10^{ 4}$ \wn\\
    $a_{3}$ &   1.94991293296139629$\times 10^{ 4}$ \wn\\
    $a_{4}$ &   3.72586955091583732$\times 10^{ 4}$ \wn\\
    $a_{5}$ &  -9.75704426363519306$\times 10^{ 4}$ \wn\\
    $a_{6}$ &  -5.91274438598366105$\times 10^{ 5}$ \wn\\
    $a_{7}$ &   1.36914591517522279$\times 10^{ 6}$ \wn\\
    $a_{8}$ &   7.68909181033474952$\times 10^{ 6}$ \wn\\
    $a_{9}$ &  -3.10832622044021636$\times 10^{ 7}$ \wn\\
   $a_{10}$ &  -8.42272589382019341$\times 10^{ 7}$ \wn\\
   $a_{11}$ &   1.75767721601389617$\times 10^{ 8}$ \wn\\
\hline
   \multicolumn{2}{c}{$ r > r_\mathrm{LR}$}\\
\hline
 ${C_3}^{\ast\ast}$ &    -5.79230$\times 10^{5}$ \wn\AA$^3$      \\
 ${C_{6}}^{\ast\ast}$ &  5.8070$\times 10^{7}$ \wn\AA$^6$   \\
 ${C_{8}}^{\ast\ast}$ & 1.276960$\times 10^{10}$ \wn\AA$^{8}$   \\
 ${C_{10}}^\ast$ &  2.0709677926$\times 10^{12}$ \wn\AA$^{10}$   \\
 ${C_{12}}^\ast$ & -8.9132400245$\times 10^{13}$ \wn\AA$^{12}$   \\
\hline
\hline
\end{tabular*}
\end{table}

\begin{table} 
			\vspace{-6.5pt}
\fontsize{8pt}{12pt}\selectfont
  \caption{Parameters of the cubic spline representation of the potential curve of state $^3\Pi_g$. The energy reference is the $5s+5p$ dissociation asymptote. $a_0=0.0529177\:$nm and 1$\:\textrm{hartree}=219474.63\:$\wn.  Parameters with $\ast$ are taken from \cite{Tsai2013,Marinescu1995}. The long-range expansion [Eq.~(\ref{eq:lr})] starts at
$r_\mathrm{LR}=22.0~a_0$.}
\vspace{5pt}
	\begin{tabular}{cccccc}
\toprule
\toprule
 $r$ ($a_0$)$\:$ & $\:$$U$ (hartree)$\:$ & $\:$$r$ ($a_0$)$\:$ & $\:$$U$ (hartree)$\:$ & $\:$$r$ ($a_0$)$\:$ & $\:$$U$ (hartree)\\ \midrule
 5.200$\:$ & $\:$0.0414574746$\:$  &  $\:$7.800$\:$ & $\:$ 0.0050294711$\:$ &  $\:$10.400$\:$ & $\:$0.0015099999   \\
 5.300 &0.0395061844  &  7.900 &  0.0044800615 &  10.500 &0.0015687878   \\
 5.400 &0.0375540883  &  8.000 &  0.0039844998 &  10.800 &0.0017400056  \\
 5.500 &0.0356091855  &  8.100 &  0.0035399820 &  10.900 &0.0017929458    \\
 5.600 &0.0336792690  &  8.200 &  0.0031436834 &  11.000 &0.0018428804    \\
 5.700 &0.0317719246  &  8.300 &  0.0027927844 &  11.100 &0.0018894054    \\
 5.800 &0.0298944546  &  8.400 &  0.0024844935 &  11.200 &0.0019322100    \\
 5.900 &0.0280538389  &  8.500 &  0.0022160467 &  11.300 &0.0019710696    \\
 6.000 &0.0262566226  &  8.600 &  0.0019847371 &  11.400 &0.0020058472   \\
 6.100 &0.0245088389  &  8.700 &  0.0017879069 &  11.500 &0.0020364700   \\
 6.200 &0.0228159252  &  8.800 &  0.0016229664 &  11.600 &0.0020629372  \\
 6.300 &0.0211826578  &  8.900 &  0.0014873949 &  11.700 &0.0020852962   \\
 6.400 &0.0196131052  &  9.000 &  0.0013787468 &  11.800 &0.0021036449   \\
 6.500 &0.0181106060  &  9.100 &  0.0012946491 &  11.900 &0.0021181152   \\
 6.600 &0.0166777722   & 9.200  & 0.0012328142 &  12.000 &0.0021288682    \\
 6.700 &0.0153164958   & 9.300  & 0.0011910310 &  13.000 &0.0020769465    \\
 6.800 &0.0140279873   & 9.400  & 0.0011671733 &  14.000 &0.0018681504    \\
 6.900 &0.0128128209   & 9.500  & 0.0011592003 &  15.000 &0.0016243556    \\
 7.000 &0.0116709855  &   9.600 & 0.0011651567 &  16.000 &0.0013942477    \\
 7.100 &0.0106019458  &   9.700 & 0.0011831760 &  17.000 &0.0011912939    \\
 7.200 &0.0096047027  &   9.800 & 0.0012114816 &  18.000 &0.0010160425    \\
 7.300 &0.0086778516  &   9.900 & 0.0012483942 &  19.000 &0.0008653122    \\
 7.400 &0.0078196483  &  10.000 & 0.0012923203 &  20.000 &0.0007354550    \\
 7.500 &0.0070280633  &  10.100 & 0.0013417780 &  21.000 &0.0006233255    \\
 7.600 &0.0063008274  &  10.200 & 0.0013953785 &  22.000 &0.0005263669    \\
 7.700 &0.0056354939  &  10.300 & 0.0014518439 &  &        \\
\midrule
\multicolumn{6}{c}{long-range parameters}\\
\midrule
\multicolumn{6}{c}{
{${C_3}^\ast=-8.905$ hartree$\cdot a_0^3 $}}\\
\multicolumn{6}{c}{
{${C_6}^\ast=8047.0$ hartree$\cdot a_0^6 $}}\\
\multicolumn{6}{c}{
{${C_8}^\ast=1133000.0$ hartree$\cdot a_0^8 $}}\\  
\bottomrule
\bottomrule
\end{tabular}
\label{pot4}
\end{table}

\clearpage

\end{document}